\documentclass{article}

\usepackage{amsmath,amssymb}

\usepackage{changepage}

\usepackage{comment}
\usepackage[utf8x]{inputenc}

\usepackage{textcomp,marvosym}

\usepackage{tabu}
\usepackage{algorithmic}
\usepackage{algorithm}
\usepackage{subfig}
\usepackage{multirow}

\usepackage{cite}

\usepackage{microtype}

\usepackage{array}

\newcolumntype{+}{!{\vrule width 2pt}}

\newlength\savedwidth

\raggedright
\usepackage{booktabs}
\usepackage[aboveskip=1pt,labelfont=bf,labelsep=period,justification=raggedright,singlelinecheck=off]{caption}

\makeatletter
\renewcommand{\@biblabel}[1]{\quad#1.}
\makeatother

\date{}

\usepackage{lastpage,fancyhdr,graphicx}
\usepackage{epstopdf}
\fancyheadoffset[L]{2.25in}
\fancyfootoffset[L]{2.25in}

\graphicspath{{./PNG/}} 

\begin{document}
\begin{flushleft}
	{\Large
		\textbf\newline{Scalable preprocessing of high volume bird acoustic data} 
	}
	\newline
	\\
	Alexander Brown,
	Saurabh Garg,
	James Montgormery
	\\
	\bigskip
	\textbf School of Technology, Environments and Design, University of Tasmania, Hobart, Tasmania, Australia
	\\
	\bigskip
	
\end{flushleft}
\section*{Abstract}
	


	In this work, we examine the problem of efficiently preprocessing high volume bird acoustic data. 
	We combine several existing preprocessing steps including noise reduction approaches into a single efficient pipeline by examining each process individually.
	 
	We then utilise a distributed computing architecture to improve execution time. Using a master-slave model with data parallelisation, we developed a near-linear automated scalable system, capable of preprocessing bird acoustic recordings 21.76 times faster with 32 cores over 8 virtual machines, compared to a serial process.
	
	This work contributes to the research area of bioacoustic analysis, which is currently very active because of its potential to monitor animals quickly at low cost. Overcoming noise interference is a significant challenge in many bioacoustic studies, and the volume of data in these studies is increasing. Our work makes large scale bird acoustic analyses more feasible by parallelising important bird acoustic processing tasks to significantly reduce execution times.


\section*{Introduction}
\label{sec:intro}

Bird monitoring has recently been of great research interest because of its broad range of applications including tracking migration\cite{Stepanian_2016}, monitoring biodiversity\cite{Salamon_2016}, and tracking population size\cite{Bardeli_2010}. Monitoring is highly important because it can be used to measure human impact on the environment\cite{Aide_2013,Bardeli_2010}. A current approach to bird monitoring is to set up sensors to record vocalisations. The science of analysing animal vocalisations is called bioacoustics. 
Almost all bioacoustic analyses require audio to be preprocessed to get it into a form suitable for analysis. This could include data compression techniques to speed up processing such as removing unnecessary audio channels \cite{Xie_2015} and downsampling\cite{Digby_2013}. It can also include improving the quality of audio by reducing noise interference, which is a key challenge for many bioacoustic studies because noise can mask vocalisations of interest\cite{Alonso_2017}. 
Noise can be considered to be any sound that is not produced by a bird. It is of great interest to remove these noises so that further processing (e.g. bird identification) can focus on the parts of a recording containing bird sound without interference. Many approaches already exist for detecting and removing noise from multiple sources\cite{Bedoya_2017,Ephraim_1984,Ferroudj_2015,Towsey_2014a,Priyadarshani_2016}.

Currently, many bioacoustic preprocessing approaches are applied individually in a manual or semi-automated way. However, such approaches are not well suited to large scale studies because of the time required to process recordings\cite{Alonso_2017,Bedoya_2014,Towsey_2014b}. Recorders are being deployed in larger numbers across different natural environments, and so are collecting bioacoustic data at high volumes, sometimes on the order of hundreds of gigabytes per day\cite{Truskinger_2014}. Moreover, preprocessing is made up of multiple steps, and previous work does not consider how to efficiently combine processes together. Thus, it is not trivial to even apply an off-the-shelf solution such as Hadoop to process such large amounts of data. While there have been some attempts to scale the processing of bioacoustic data using distributed systems\cite{Dugan_2011,Thudumu_2016,Truskinger_2014}, these do not focus on preprocessing steps, and use off-the-shelf solutions (e.g. Hadoop, or MATLAB distributed file system) which add overhead and do not utilise low level control over data, resulting in inefficiencies. 





In this work, we examine how to preprocess high volume bird acoustic data quickly and efficiently. To achieve this, we combine existing preprocessing steps into an efficient processing pipeline. This includes compression and the removal of several types of noise, namely stationary background noise, rain, and cicada choruses.  We also remove silence to improve processing efficiency. The order in which to perform these approaches is significant in that time can be saved by skipping unneeded processes in some files. This order is determined here by examining how much audio each process removes, and the effect of some processes on the accuracy of others.

To increase processing speed, we derive a mechanism to distribute this unified pipeline across multiple machines in an efficient and scalable manner. This greatly increases the computing power available for processing the pipeline, increasing processing speeds and making the processing of very large amounts of bioacoustics data more feasible. An emphasis is placed on scalability, aiming for linear proportionality between the improvement rate of execution time and the amount of resources used.

\section*{Background}
\label{sec:overview}

This section introduces the objectives of our work, before listing the processes we will be using. It then introduces the processing pipeline, giving a brief overview of how we derive it, before discussing how we approach the distribution of the pipeline.

\subsection*{Pipeline processes}
\label{sec:nrProcesses}

This work focuses on improving the efficiency of preprocessing bird acoustic recordings, which can later be used for further analysis, such as species detection.  
The preprocessing stage consists of the following tasks:

\begin{itemize}
	\item \textbf{Splitting}: Audio is split into smaller chunks which allows for work to be distributed more easily. Additionally, long files are not viable for processing on their own because of high RAM requirements\cite{Truskinger_2014}, and some classification tasks in the pipeline work better on shorter samples.
	\item \textbf{Downsampling}: Audio files have sample rates converted to 22.05 kHz to reduce their size. Bird sounds are normally below 11.025 kHz (the Nyquist frequency)\cite{Pijanowski_2011}, so signals of interest are not lost.
	\item \textbf{Converting to Mono}: Only one channel of audio is needed to detect significant audio signals, so this is used to further reduce the size of files.
	\item \textbf{High-Pass Filter (1 kHz)}: Birds typically do not emit sound below 1 kHz\cite{Pijanowski_2011}, so all data below this frequency is noise and hence is attenuated.
	\item \textbf{Sound Enhancement}: Stationary background noise is reduced. While there are several approaches that can achieve this\cite{Boll_1979,Ren_2008,Priyadarshani_2016} we use the Minimum Mean Square Error Short Time Spectral Amplitude estimator (MMSE STSA) filter\cite{Ephraim_1984}, which was found in separate work \cite{Alonso_2017} to be highly effective.
	\item \textbf{Short-Time Fourier Transform}: Time-based information is transformed into frequency-based information. Several acoustic indices used in cicada and rain 
	detection use frequency-based information, so this is only executed once, rather than for each acoustic index calculated, or for each process. The FFT implementation used here is from the Apache Commons Math library\cite{Apache_Math} and is described by Demmel\cite{Demmel_1997}. A window size of 256 samples is used with Hamming windows with 50\% overlap.


	\item \textbf{Heavy Rain Detection and Removal}: Heavy rain is detected by using rules derived from a C4.5 classifier\cite{Quinlan_1993} using acoustic indices. This approach is similar to Towsey et al.\cite{Towsey_2014a} and Ferroudj\cite{Ferroudj_2015}. Spectral-based signal to noise ratio and power spectral density used by Bedoya et al.\cite{Bedoya_2017} were added to the acoustic indices used in the classifier. The classifier was trained on a separate sample of data and its rules then hard coded into our Java-based implementation prior to beginning the pipeline.
	\item \textbf{Cicada Detection}: Cicada choruses are detected using the same general approach as rain detection.
	\item \textbf{Cicada Removal}: Cicada choruses are removed using band-pass filters to eliminate audio from frequency ranges containing cicada choruses. These ranges are calculated by examining FFT coefficients. 
\end{itemize}

\subsection*{Problem objectives}

This work aims to improve the speed and efficiency of preprocessing bird acoustic data by combining existing preprocessing tasks into an efficient pipeline and applying this pipeline in a distributed system. This is done so that large data sets can be processed in a reasonable time, which is becoming increasingly important because of increasing amounts of data being recorded\cite{Truskinger_2014}. In this work, we do not focus on improving the efficiency of individual preprocessing tasks.

\subsection*{Challenges}

\subsubsection*{Unification of processes}

The first key challenge in achieving the research objectives is to determine an efficient approach to compose noise removal processes as a single system. This requires several questions to be answered, such as whether different sequences of the denoising tasks affect their accuracy and whether executing some tasks earlier can improve the overall efficiency of the pipeline. In other words, we need to investigate the trade-off between two factors: execution time of each process and how they influence the accuracy of each other when applied in a pipeline. We also consider which lengths of audio are best for performing denoising in terms of both accuracy and execution time.


\subsubsection*{Distribution of tasks for large scale processing}
\label{sec:distOverview}

To support the pre-processing of large volume bird acoustic data distributed computing approaches can be utilised. However, determining which approach should be employed for this problem of scalable processing still needs to be investigated. For this research, we aim to achieve linear scalability. This means that improvements in execution time (i.e. in terms of ratios) are linearly proportional to the number of processors used. 
\begin{itemize}
\item\textbf{Distributed Computing Architectures:}
There are typically two distributed system architectures in literature: the master-slave and peer-to-peer models \cite{Krauter_2002}. These determine how different components of the system communicate with each other, and also guide how work is distributed. In the master-slave model, a single master process manages multiple identical slave processes and distributes work to them. This approach is simpler than many other models, such as peer-to-peer, because the master process handles all work distribution. However, it is also less fault tolerant than other architectures, as the master process is a central point of failure, and can be less scalable than other approaches if, for example, the master process is being overworked, creating a bottleneck\cite{Krauter_2002}.

Another architecture is the peer-to-peer model. This is the opposite of the master-slave model, where workload is decentralised. Because of the decentralisation, peer-to-peer networks are more adaptable than master-slave networks, and can be highly scalable\cite{Androutsellis_2004}. However, this model is also more complex to work with in many cases, because communication can occur between any two systems in the network, which can create extra overhead and ultimately slow the system down\cite{Krauter_2002}.

As such, a master-slave model is well suited for the present system, as different parts of the audio can be preprocessed independently without any requirement of communication. The master can simply split files, and distribute them to slaves. This should be scalable, because the master does not perform much work in splitting audio files and managing distribution, relative to the overall pipeline. This approach is comparable to other work with large scale bioacoustics analyses\cite{Dugan_2011,Thudumu_2016,Truskinger_2014}.


\item\textbf{Parallelisation Approaches:} In addition to deciding which architecture to use for our system, we must also consider how to parallelise the workload. Here, we will examine two such approaches: data parallelisation and functional parallelisation.

Data parallelisation involves dividing data between machines, and having each machine apply processing on the data it receives. This is most well suited to problems where data can be easily split and divided evenly, and processed independently. Functional parallelisation involves having machines process different functions on the same data. This allows for multiple processors to work on the same data in parallel, but is more difficult to evenly distribute work, particularly if different functions take different amounts of time to execute.

Data parallelisation is well suited to the pre-processing of bioacoustics data. The nature of audio recordings makes them easy to divide into small chunks, and have each chunk processed on a different machine. Furthermore, detection processes require files to be split into small chunks anyway (e.g., it makes little sense to decide if a single day-long sample is silent). Additionally, processes in our pipeline execute at very different speeds, and some can remove audio without completing subsequent steps, complicating a potential functional parallelisation approach.

\end{itemize}

While we could use an off-the-shelf system such as Hadoop\cite{Shvachko_2010} or Spark\cite{SparkStreaming} to achieve this parallelisation, these do not give low level control over data in order to maximise efficiency. A previous attempt to utilise Hadoop and Spark for some preprocessing steps (such as splitting bioacoustic audio files and generating spectrograms) by Thudumu et al.\cite{Thudumu_2016} did not achieve linear scalability. Moreover, for the best results, the investigation of the exact split length of each audio file for each pre-processing task, the sequence of each task and how they are distributed for linear scalability are still needed. Therefore, in this paper we investigate these factors and propose a master-slave based data parallelisation system for pre-processing high volume bioacoustic data.

\section*{Processing pipeline}
\label{sec:pipeline}

The processing pipeline unifies preprocessing tasks previously described to prepare bioacoustic data for future analysis. We aim to do this as efficiently as possible, while maintaining the accuracy of detection processes. As such, the execution order is important, because some processes remove or modify the audio. Removed audio does not need to be processed by subsequent steps in the pipeline, increasing efficiency, whereas modified audio affects the accuracy of subsequent detection steps, which affects the overall effectiveness of the pipeline.

The pipeline is derived by first evaluating execution times for each process, and how this varies with the lengths of audio chunks processed at a time. We then evaluate the accuracy of noise detection processes before and after applying the MMSE STSA filter, and finally test to see if detection approaches have a dependency on split length.

\subsection*{Evaluation for sequencing of the processing pipeline}


Three experiments are conducted to help in developing the processing pipeline. The first experiment looks at the computation times for each processing step, and how these vary depending on the size of data they are processing at once (called file split size/length). This experiment identifies fast and slow processes. Faster processes are placed earlier in the pipeline where possible if they can result in later, slower processes being skipped for some data (i.e. due to the deletion of audio). This experiment can help to identify which split lengths result in faster execution for each process, which can be used to improve their execution time.

The second experiment examines the effect of the MMSE STSA filter, which alters audio files in a significant way and affects detection processes. As such, we test the accuracy of detection approaches before and after applying the filter.

The final experiment looks at whether detection accuracy is dependent on the length of chunks into which the audio is split. We take a random 30 minute sample extracted from four days of unsupervised environmental recordings, manually classify rain and cicada choruses and compare this to the automatic classifiers. This can show if detectors work better on certain lengths. This is important in determining the processing order, because files can only be split, and not joined (as adjacent chunks may be sent to different slaves), meaning detection processes with longer split lengths will need to run earlier than those that do not.

\subsubsection*{Recording data}

Environmental recordings for evaluating the system have been provided by the Samford Ecological Research Facility (SERF), based in the Queensland University of Technology (QUT). These recordings were taken over five days between 12 October 2010 and 16 October 2010, over four sensors, for a total of 20 days of audio to process. In practice, four days of recordings are used in testing. Recordings from this group have been used in several studies before\cite{Towsey_2014a,Towsey_2014b,Truskinger_2014}. While these recordings are of high quality, they do contain significant levels of background noise, large variations in the loudness of bird sounds, ranging from very clear to barely audible, and noise interference from many sources including rain and cicadas, which makes the sample well suited for this study.

\subsubsection*{Per-step execution time}
\label{sec:perstep}
A test is conducted where each step is performed independently. Two hours of audio known to contain rain, 
cicada choruses and bird sounds is passed through the processing pipeline in sequence, using one processor. The split length is varied (from 5 to 30 seconds in 5-second increments) to observe its effects on processing time. Each test is completed five times for each split length, and the average and standard deviation of the computation times are taken.

\begin{figure}
	\centering
	\includegraphics[width=1\linewidth]{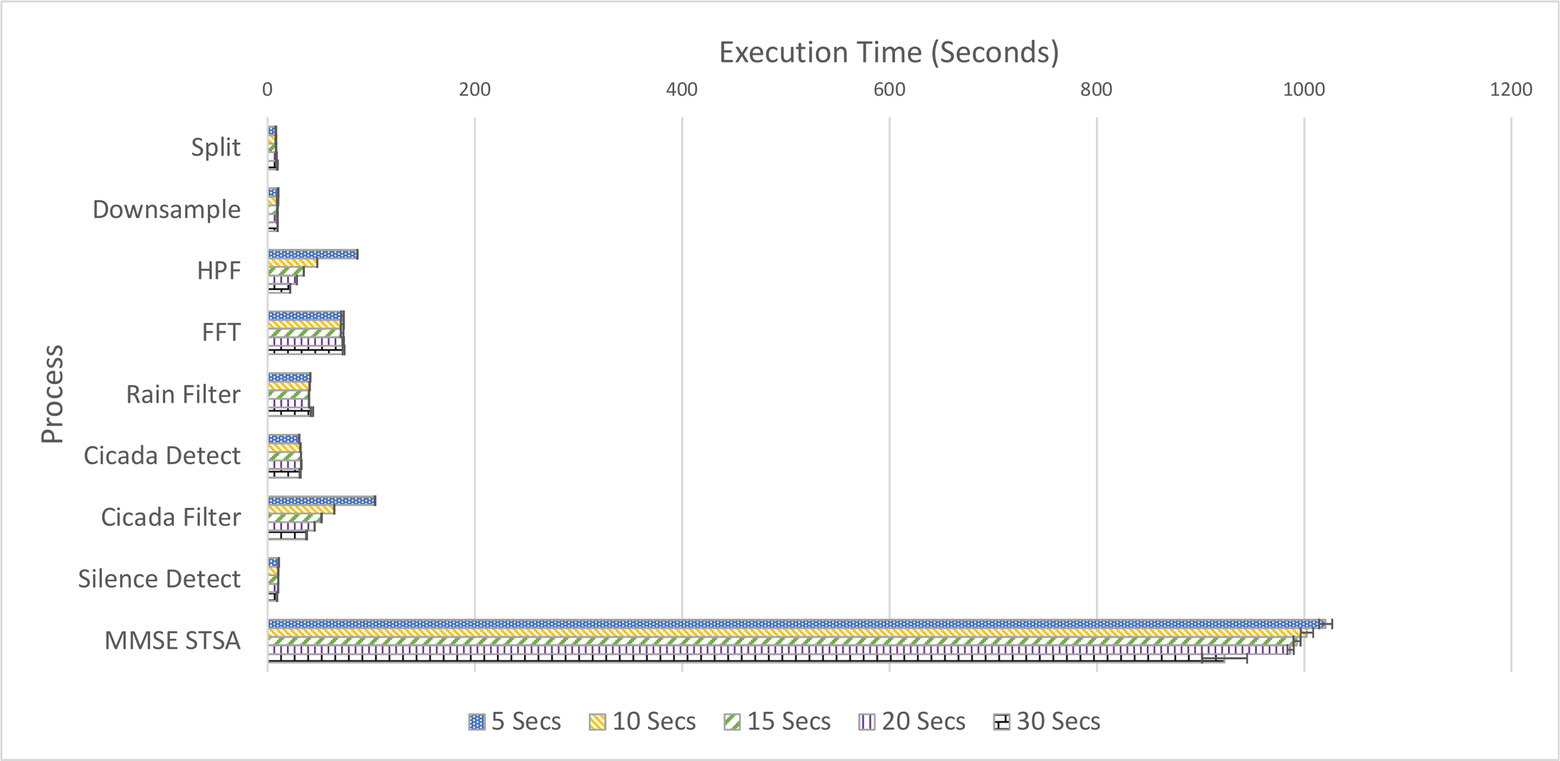}
	\caption{\textbf{Computation times per process for different split lengths up to cicada detection.} Error bars indicate standard deviation (FFT = Fast Fourier Transform, HPF = High-Pass Filter, MMSE STSA = Minimum Mean Square Error Short Time Spectral Amplitude filter)}
	\label{Fig.:processtimesperprocess}
\end{figure}

Fig.~\ref{Fig.:processtimesperprocess} and Table~\ref{tab:compTimes} shows the execution times for all processes for 2 hours (1.2 GB) of audio. Each process is applied to every file, although, once the pipeline is developed, not all processes are applied to every file, as some files may be removed because they contain rain
.

\begin{table*}
	\caption{Computation times for each processing step in relation to split lengths with standard deviations}
	\footnotesize
	\centering
	\begin{tabu} to 1\textwidth{X[2,c]X[1,c]X[1,c]X[1,c]X[1,c]X[1,c]}
		\toprule
		\multirow{2}{*}{Processing Step} &\multicolumn{5}{c}{Split Length (seconds)}
		\\
		&5&10&15&20&30
		\\
		\midrule
		Splitting&7.85$\pm$0.42&7.95$\pm$0.49&8.13$\pm$0.51&9.24$\pm$0.42&8.87$\pm$0.42
		\\
		\midrule
		Downsampling & 10.18$\pm$0.42&9.59$\pm$0.68&9.30$\pm$0.30&9.29$\pm$0.52&9.57$\pm$0.19\\
		\midrule
		High-pass Filter & 86.63$\pm$0.13&47.79$\pm$0.17&34.8$\pm$0.18&28.2$\pm$0.11&21.67$\pm$0.09\\
		\midrule
		Fast Fourier transform&2.39$\pm$1.01&47.79$\pm$1.44&71.90$\pm$1.36&73.15$\pm$0.56&73.21$\pm$0.95 \\
		\midrule
		Rain Filter&41.11$\pm$0.20&40.46$\pm$0.20&39.86$\pm$0.15&39.94$\pm$0.18&42.67$\pm$1.16\\
		\midrule
		Cicada Detection & 30.47$\pm$0.20&31.58$\pm$0.20&32.04$\pm$0.08&32.32$\pm$0.26&31.36$\pm$0.60\\
		\midrule
		Cicada Filter &103.48$\pm$0.56&64.30$\pm$0.18&51.94$\pm$0.22&45.27$\pm$0.23&37.46$\pm$0.52\\
		\midrule
		MMSE STSA & 1020.57$\pm$6.49&1002.65$\pm$5.98&993.10$\pm$3.39&986.92$\pm$3.09&923.21$\pm$21.78
		\\
		\bottomrule
	\end{tabu}
	\label{tab:compTimes}
\end{table*}

The figure shows two distinctive features. First is the large decrease in the execution time of the high-pass, cicada, and MMSE STSA filters filters when the split size is larger. The differences in high-pass and cicada execution times are likely due to the use of the non-native sound processing library Sound eXchange (SoX)\cite{SoX}. This causes extra overhead with each call, and shorter split sizes require more calls to SoX. This is more of a problem for high-pass filtering than cicada filtering, as this is executed on every file, whereas cicada filtering only applies to parts of the recording where cicada choruses are detected, which, as determined by subsequent testing, is a small fraction of the total recording.

The second observation is that the MMSE STSA filter takes much longer than the other processing steps combined. As such, execution time can be significantly saved by removing audio before the MMSE STSA filter is applied.

\begin{figure}
	\centering
	\includegraphics[width=0.7\linewidth]{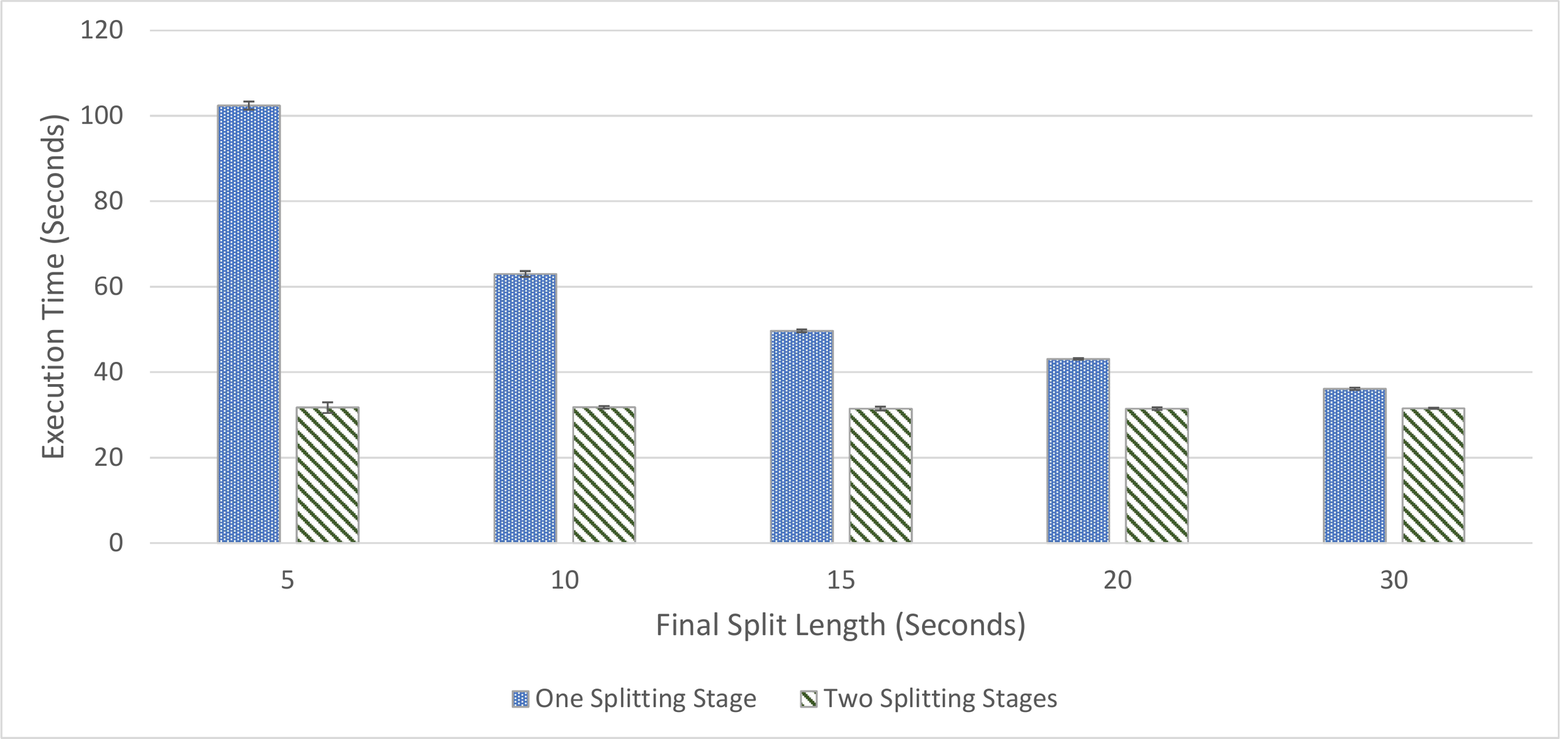}
	\caption{\textbf{High-pass filtering computation times} High-pass filtering computation times comparison, between splitting to the final length, downsampling, and then high-pass filtering (one split) and splitting to 1-minute (2.5 MB) chunks first, downsampling and high-pass filtering, then splitting to the final length (two splits)}
	\label{Fig.:hpftimes}
\end{figure}

The trend in high pass filter execution time gives rise to a potential improvement. If clips are split into larger chunks first, downsampled and high-pass filtered, and then split into smaller chunks, execution time can be improved. Testing an approach that performs this shows an improvement in execution time, as shown in Fig.~\ref{Fig.:hpftimes}. Here, audio is split into 1-minute (2.5 MB) long chunks, downsampled, high-pass filtered, and then split to the target split length. Two hours of audio is tested against two approaches, one that splits audio to the target length immediately, and one that split files into 1-minute long chunks first, and then splits again.

While it would be theoretically optimal to run a high-pass filter on whole audio files, rather than running an initial split to 1-minute long chunks, some consideration needs to be made for when this pipeline is processed in parallel, where it is advantageous to start allocating files to machines to process as quickly as possible, and to give them shorter files such that work can be distributed more evenly. As such, this initial split length is used as an input parameter to test the distributed system to find an efficient configuration.

\subsubsection*{Silence removal}

As discussed above, 
 it is highly advantageous to remove audio before execution the MMSE STSA filter because of its long execution time. Audio containing heavy rain is already removed, but even more audio can be removed by detecting audio that does not contain any bird sound of interest. Because of this, we introduce a basic silence removal approach to the processing pipeline. This approach uses a simple threshold. The choice of threshold is derived next, 
 based on one of two acoustic indices taken from Bedoya et al.\cite{Bedoya_2017}: Power Spectral Density (PSD), and Signal to Noise Ratio (SNR). In testing the execution time of this silence detection approach, we found it takes a very short time relative to other processes, taking approximately 10 seconds to process 2 hours (1.2 GB) of audio, regardless of the split length.

Silence detection testing is now added to subsequent tests used in evaluating the processing pipeline.

\subsubsection*{Effect of the MMSE STSA filter on noise reduction}
\label{sec:mmse}

The Minimum Mean Square Error Short Time Spectral Amplitude estimator (MMSE STSA)\cite{Ephraim_1984} is a process within the processing pipeline that reduces stationary background noise. Because this process makes signals clearer, it seems likely that this process would improve the accuracy of detection processes. However, this process is time consuming, as shown in Fig.~\ref{Fig.:processtimesperprocess}, so processes should only be applied after the MMSE STSA filter if they show significant improvement in detection accuracy, particularly if these processes remove audio, as removed audio does not need to be processed further. 
Here, we test the accuracy of rain, cicada, and silence filters before and after applying the MMSE STSA filter to determine where they belong in the pipeline, relative to the MMSE STSA filter. 

We first evaluate the accuracy of rain and cicada detection when the MMSE STSA filter is applied. For this test, acoustic indices were calculated for raw audio, and audio processed by the MMSE STSA filter (although a 1 kHz high-pass filter was used for each set). The audio in each set was otherwise identical outside of processing. 

The classification accuracies of each set are given in Table~\ref{tab:MMSEDetectAccuracy}. This clearly shows that the MMSE STSA filter does not improve accuracy, and actually reduces it for rain detection. This is likely because rain has stationary and non-stationary components (i.e. raindrops distant from the sensor make a constant background noise, whereas closer raindrops are clearly audible and distinguishable). As such, the MMSE STSA reduces some, but not all of the noise sources, making them more difficult to detect.


%
\begin{table}
	\caption{Comparison of Detection Accuracy Depending on Use of MMSE STSA Filter}
	\label{tab:MMSEDetectAccuracy}
	\centering
	\begin{tabular}{lcc} 
	\toprule
	Filter & Cicada Accuracy & Rain Accuracy\\
	\midrule
	Raw & 99.3\% & 96.9\% \\ 
	MMSE STSA & 99.1\% & 92.9\% \\
	\bottomrule
	\end{tabular}
\end{table}

\begin{figure}
	\centering
	\includegraphics[width=0.8\linewidth]{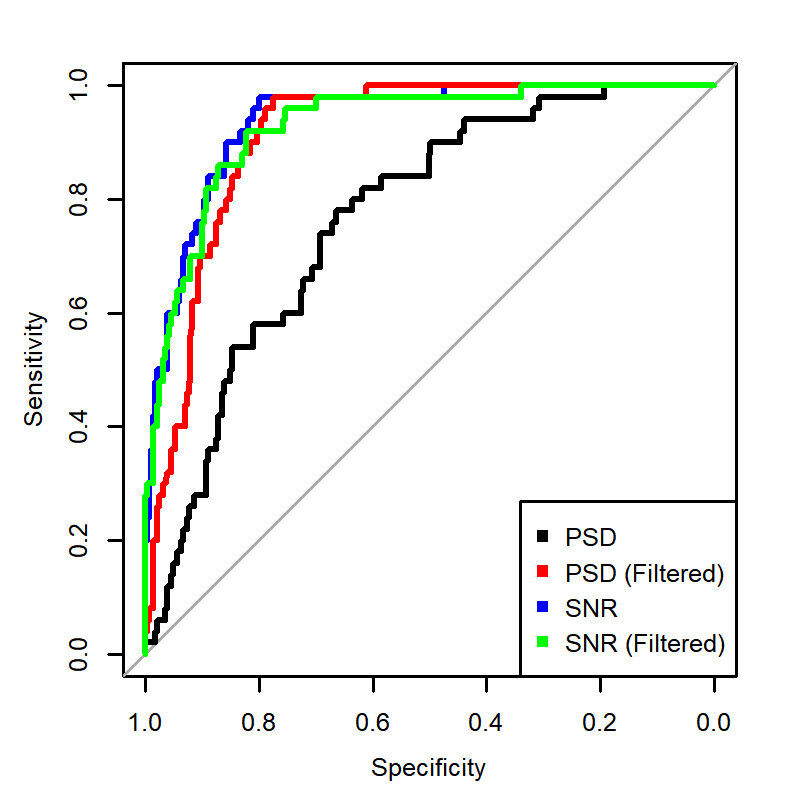}
	\caption{\textbf{ROC Curve for Classifying Silence}}
	\label{Fig.:roccurvesilence}
\end{figure}

For silence detection, thresholds using two different measures were considered: power spectral density and signal to noise ratio (SNR). These were applied to files with and without the MMSE STSA filter to evaluate accuracy. Because only one measure is used, an ROC curve (Fig.~\ref{Fig.:roccurvesilence}) was employed to visualise the accuracy of the thresholds as they were increased, in terms of the sensitivity and selectivity. The Area Under the Curve (AUC) was taken for each threshold and recording set, shown in Table~\ref{tab:AUCSilence}.

\begin{table}
	\caption{Area Under the Curve (AUC) results for Silence Removal, with 95\% Confidence Intervals (CI) for raw and MMSE STSA filtered audio using Power Spectral Density (PSD) and Signal to Noise Ratio (SNR) thresholds.}
	\centering
	\begin{tabular}{llcc}
		\toprule
		Audio Source & Index & AUC & 95\% CI\\
		\midrule
		Raw & PSD & 0.768 & 0.745--0.831\\ 
		Raw & SNR & 0.939 & 0.910--0.969\\
		Filtered & PSD & 0.913 & 0.8818--0.944 \\
		Filtered & SNR & 0.929 & 0.894--0.964 \\
		\bottomrule	
	\end{tabular}
	\label{tab:AUCSilence}
\end{table}

The results of this show that, if using the Power Spectral Density measure, the MMSE STSA filter would be necessary to obtain good results. However, the SNR measure performs similarly well regardless of the use of the MMSE STSA filter. Because of the time cost of using the MMSE STSA filter, it is more efficient to execute silence detection based on SNR prior to executing the MMSE STSA filter.


%

%

\subsubsection*{Effect of split length on noise reduction}
\label{sec:splitLengthDetection}

This section examines if detection approaches are dependent on split lengths. To do this, the accuracy of each detector (silence, rain, and cicada chorus) is tested on 30 minutes of audio composed by randomly selected 1-minute chunks spread over four days of original recordings. These chunks were then split into 5, 10, 15, 20, and 30 second chunks (these divide evenly into 60 seconds). These were listened to and manually labeled as rain, cicada, or silence, to a resolution of 5 seconds. Each detection approach was then tested for each split length. Manual labelling was performed on audio filtered by the MMSE STSA algorithm, even though automatic methods work with raw audio. This gives better accuracy for manual labelling, particularly for detecting silence, because very quiet calls become clearer.

Accuracy is evaluated for each split length to a precision of 5 seconds, despite the fact that these approaches do not have this level of precision for longer split lengths. For example, given a 10-second long chunk, if there is silence in the first 5 seconds, but a sound in the following 5 seconds, and that chunk is labelled as silence by the system, this is interpreted as one true positive and one false positive result, even though only one file was classified.

In practice, the silence classifier labels some rain as silence. This makes intuitive sense, given it is using an estimated signal to noise ratio (SNR) threshold, which is a measure of peak volume to average volume. If the average volume is very loud then the SNR will be low, even if the peak volume is also loud (compared to times when it is not raining). Despite technically being a false positive, this is not a significant issue, because rain is removed by the rain filter anyway. However, this creates a complication, because some rain samples contain audible rain drops, which results in files with a high signal to noise ratio. Consequently, because the silence filter detects some, but not all rain samples as containing silence, samples manually classified as containing rain were removed from the silence classification test. 

In all figures in this section, the number of true positives, false positives, and false negatives are shown. True negatives are excluded from these figures as the number of true negatives is much greater than the others in every case, which makes visual comparison more difficult. 

\begin{itemize}
	\item \textbf{Cicadas:} The cicada detection results, depicted in Fig.~\ref{Fig.:cicadadetection} and Table~\ref{tab:cicada_accuracy}, shows that cicada detection works well for all split lengths, detecting all cicada choruses in the sample, with a small number of false positives. The best performing split length is 15 seconds, which contained no false positives, although this strong result could be partially due to chance.

	
	\begin{figure}
		\centering
		\includegraphics[width=0.8\linewidth]{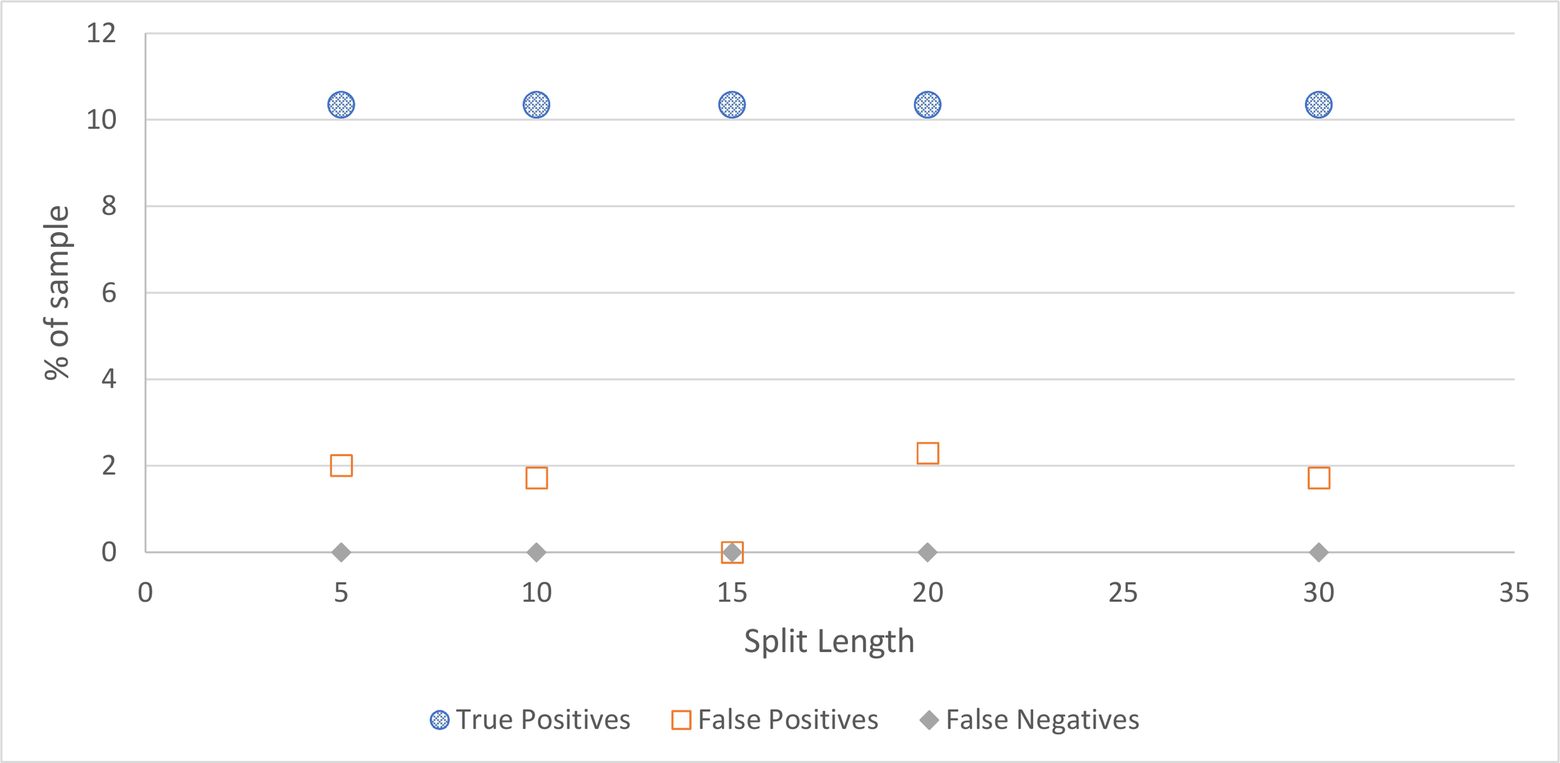}
		\caption{\textbf{Results of cicada classification test.}}
		\label{Fig.:cicadadetection}
	\end{figure}


	\begin{table}
		\centering
		\caption{Cicada detection accuracy}
		\label{tab:cicada_accuracy}
	\begin{tabular}{cccccc}
		\toprule
		Split & True & False & False & True & \\
		Length & Pos. & Pos. & Neg. & Neg. & Accuracy\\
		\midrule
		5&10.3\%&2.0\%&0.0\%&87.6\%&98.0\%\\
		10&10.3\%&1.7\%&0.0\%&87.9\%&98.3\%\\
		15&10.3\%&1.7\%&0.0\%&89.7\%&100.0\%\\
		20&10.3\%&2.3\%&0.0\%&87.4\%&97.7\%\\
		30&10.3\%&1.7\%&0.0\%&87.9\%&98.3\%\\
		\bottomrule
	\end{tabular}
	\end{table}

	\item \textbf{Rain:} Similar to cicada detection, the amount of rain detected does not vary much depending on split length, as shown in Fig.~\ref{Fig.:raindetection} and Table~\ref{tab:rain_accuracy}. Somewhat surprisingly, rain detection is slightly more sensitive, and more accurate, for longer split lengths, at least up to 30 seconds, at which point a steep drop-off occurs. This is likely because rain tends to occur over a long duration, and patterns that can be used to detect rain are clearer over longer time periods.
	
		\begin{figure}
		\centering
		\includegraphics[width=0.8\linewidth]{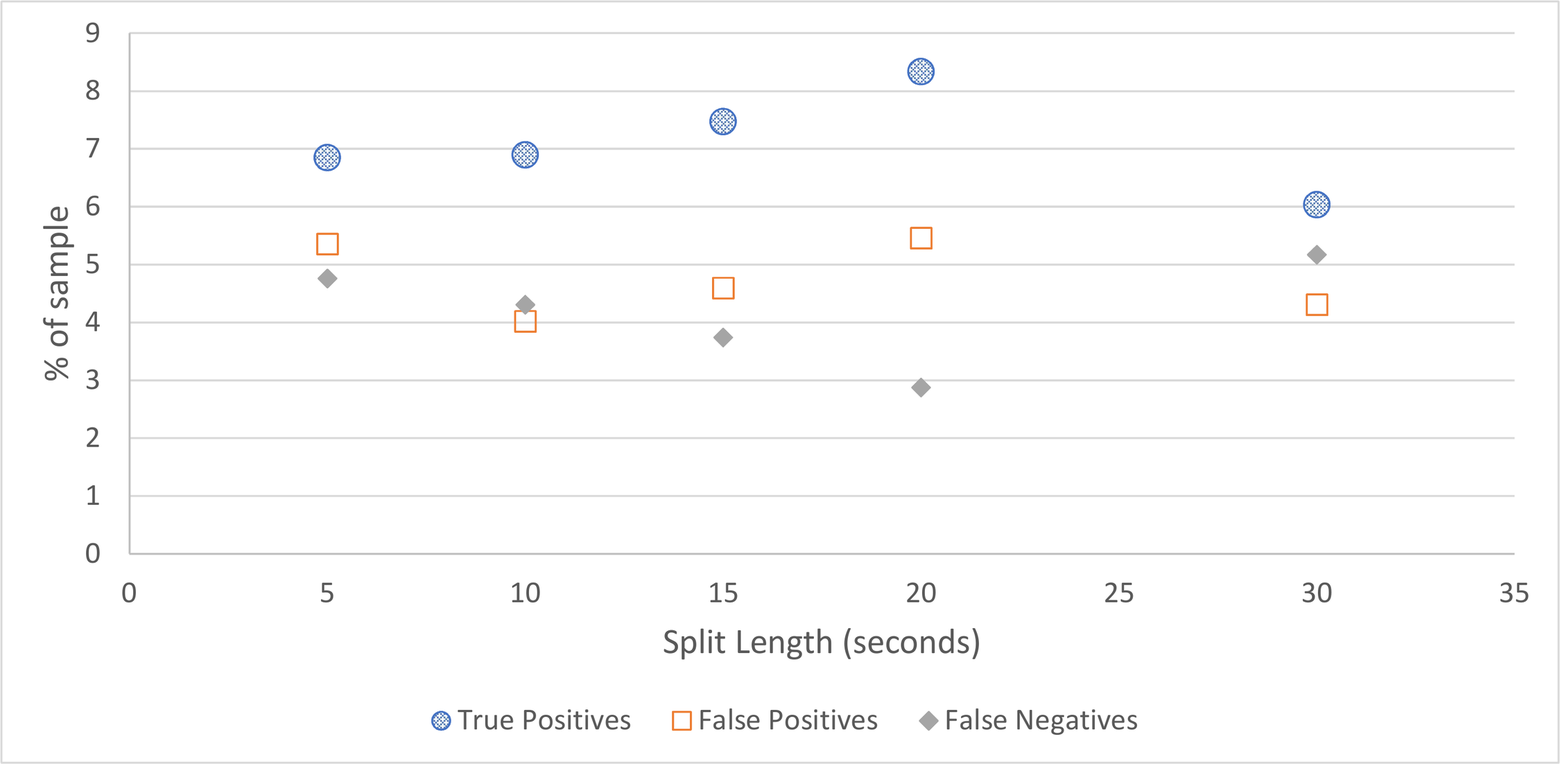}
		\caption{\textbf{Amount of audio detected as rain in a sample as it varies with split length.}}
		\label{Fig.:raindetection}
	\end{figure}

	\begin{table}
	\centering
	\caption{Rain detection accuracy}
	\label{tab:rain_accuracy}
	\begin{tabular}{cccccc}
		\toprule
		Split & True & False & False & True & \\
		Length & Pos. & Pos. & Neg. & Neg. & Accuracy\\
		\midrule
		5&6.8\%&5.4\%&4.8\&&83.0\%&89.9\%\\
		10&6.9\%&4.0\%&4.3\%&84.7\%&91.7\%\\
		15&7.5\%&4.6\%&3.7\%&84.2\%&91.7\%\\
		20&8.3\%&5.5\%&2.9\%&83.3\%&91.7\%\\
		30&6.0\%&4.3\%&5.2\%&84.5\%&90.5\%\\
		\bottomrule
	\end{tabular}
\end{table}
	
	In practice, the accuracy of rain detection is not as poor as this evaluation suggests. When manually labelling the data, only rain considered intense enough to drown out any bird signal was classified as rain, although the rain classifier classifies some lighter rain without significant bird sound as containing rain. While these are labelled as false positives, many of these would be (validly) removed by the silence detector anyway.


	\item \textbf{Silence:} Figs~\ref{Fig.:silencedetection} and~\ref{Fig.:silencedetection2}, and Table~\ref{tab:silence_accuracy}, show the effectiveness of the silence detector at different signal to noise ratio thresholds. Unlike rain and cicada detection, split length has a significant effect on the sensitivity of silence detection. This is because silence is much more likely to occur over shorter durations.
	
		\begin{figure}
		\centering
		\includegraphics[width=0.8\linewidth]{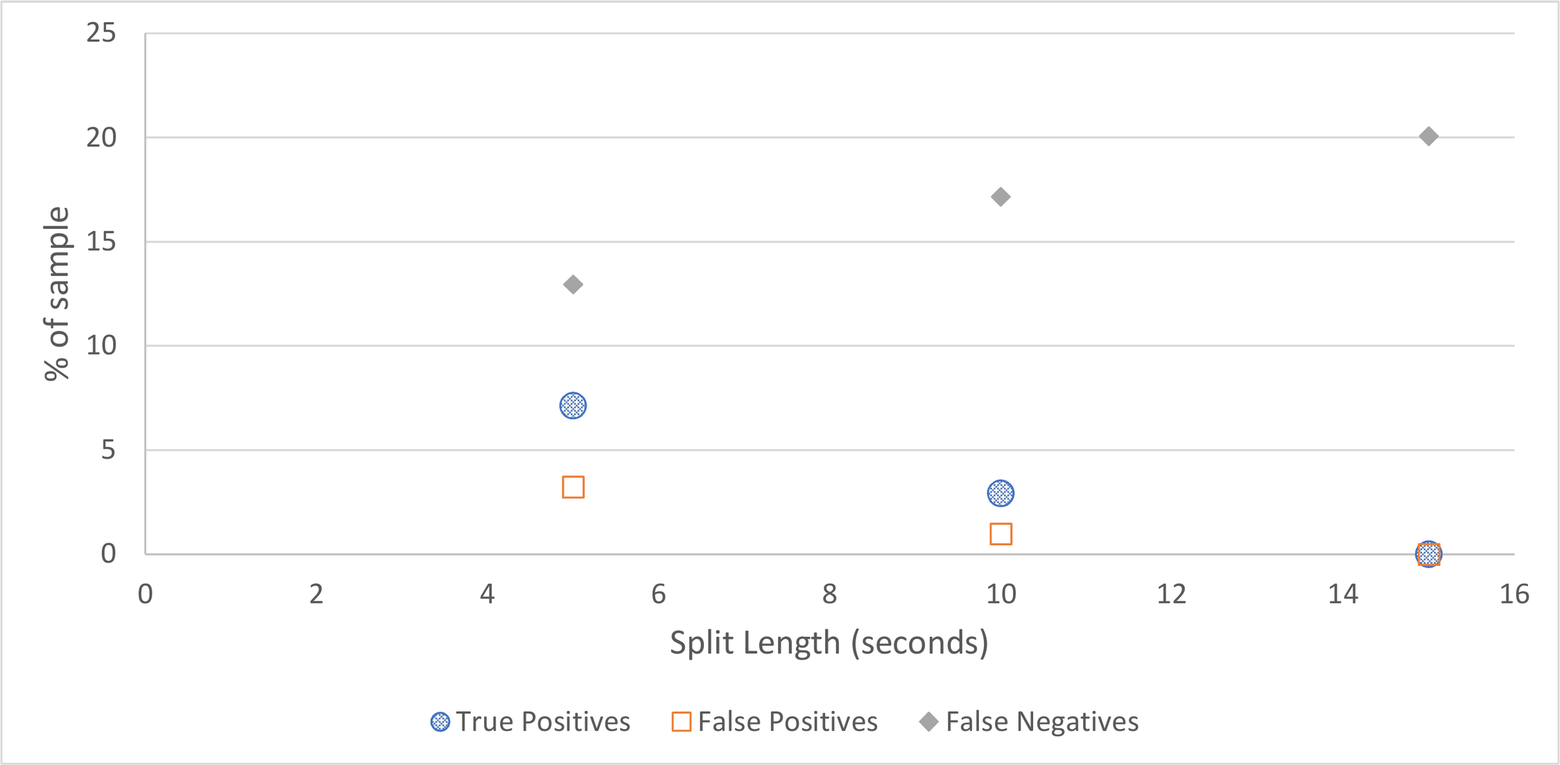}
		\caption{\textbf{Silence detection accuracy for the higher of the two thresholds tested}}
		\label{Fig.:silencedetection}
	\end{figure}
	
	\begin{figure}
		\centering
		\includegraphics[width=0.8\linewidth]{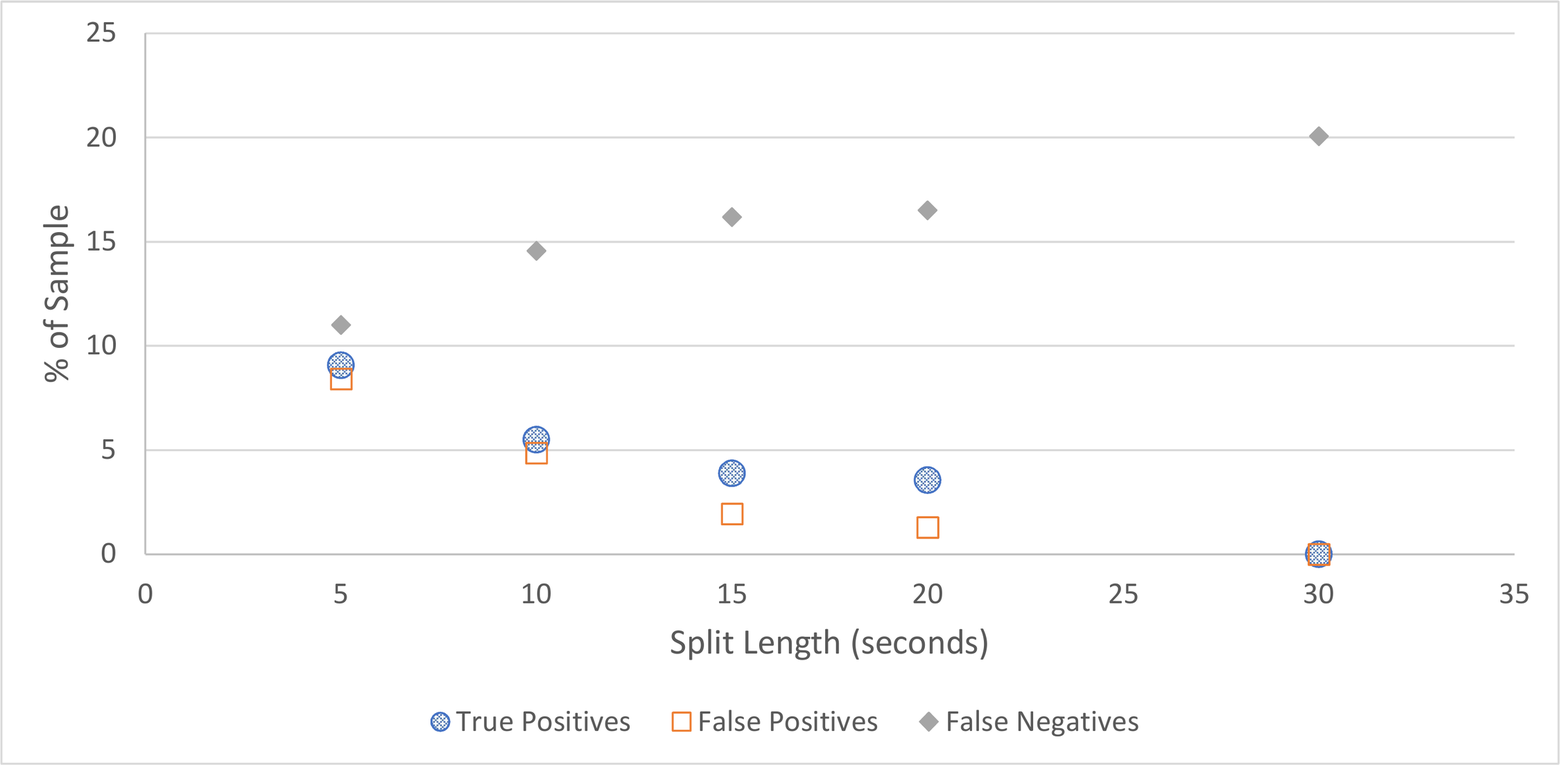}
		\caption{\textbf{Silence detection accuracy for the lower of the two thresholds tested}. All split lengths above 15 seconds detect no silence.}
		\label{Fig.:silencedetection2}
	\end{figure}

\begin{table}
	\centering
	\caption{Silence detection accuracy}
	\label{tab:silence_accuracy}
	\begin{tabular}{cccccc}
		\toprule
		Split & True & False & False & True & \\
		Length & Pos. & Pos. & Neg. & Neg. & Accuracy\\
		\midrule
		\multicolumn{6}{l}{\textit{SNR threshold = 0.25}} \\
		5&9.1\%&8.4\%&11.0\%&71.5\%&80.6\%\\
		10&5.5\%&4.9\%&14.5\%&78.0\%&80.5\%\\
		15&3.9\%&1.9\%&16.2\%&78.0\%&81.9\%\\
		20&3.6\%&1.3\%&16.5\%&78.6\%&82.2\%\\
		30&0.0\%&0.0\%&20.7\%&79.9\%&79.9\%\\
		\midrule
		\multicolumn{6}{l}{\textit{SNR threshold = 0.2}} \\
		5&7.2\%&3.3\%&12.9\%&79.9\%&83.8\%\\
		10&2.9\%&1.0\%&17.2\%&78.9\%&80.0\%\\
		15&0.0\%&0.0\%&20.1\%&79.9\%&79.9\%\\
		\bottomrule
	\end{tabular}
\end{table}
	
	Overall, the silence detector performs somewhat poorly, producing about as many false positives as true positives on more aggressive settings, and failing to detect many instances of silence on all settings, with worsening performance for longer split lengths and lower thresholds. This indicates a better approach is needed for removing silence overall, which will be the subject of future work. For the present investigation a less sensitive threshold is selected, as this is more accurate overall and retains more samples containing bird sound, which is more important than any efficiency gained from removing silence, as these can be dropped at a later point. As such, the 5-second sample with the lower threshold is considered the best setting for our filter, which does remove over one third of silence, while classifying relatively few false positives. Though using 5 second splits means that the MMSE STSA filter takes longer to execute (see Fig.~\ref{Fig.:processtimesperprocess}), the effect of removing silence will have a greater effect on reducing execution time overall.
	
	It is notable that, while the silence detector does produce many false positives, the false positives contain quiet bird calls, not significantly louder than the background noise. Even after applying the MMSE STSA filter, noise is still very audible in comparison to the bird call of interest (which consequently are poorer candidates for automated species identification anyway). In our testing, the silence filter never removed any audio with very clear bird calls.


\end{itemize}

\subsection*{Final pipeline}
\begin{figure}
		\centering
		\includegraphics{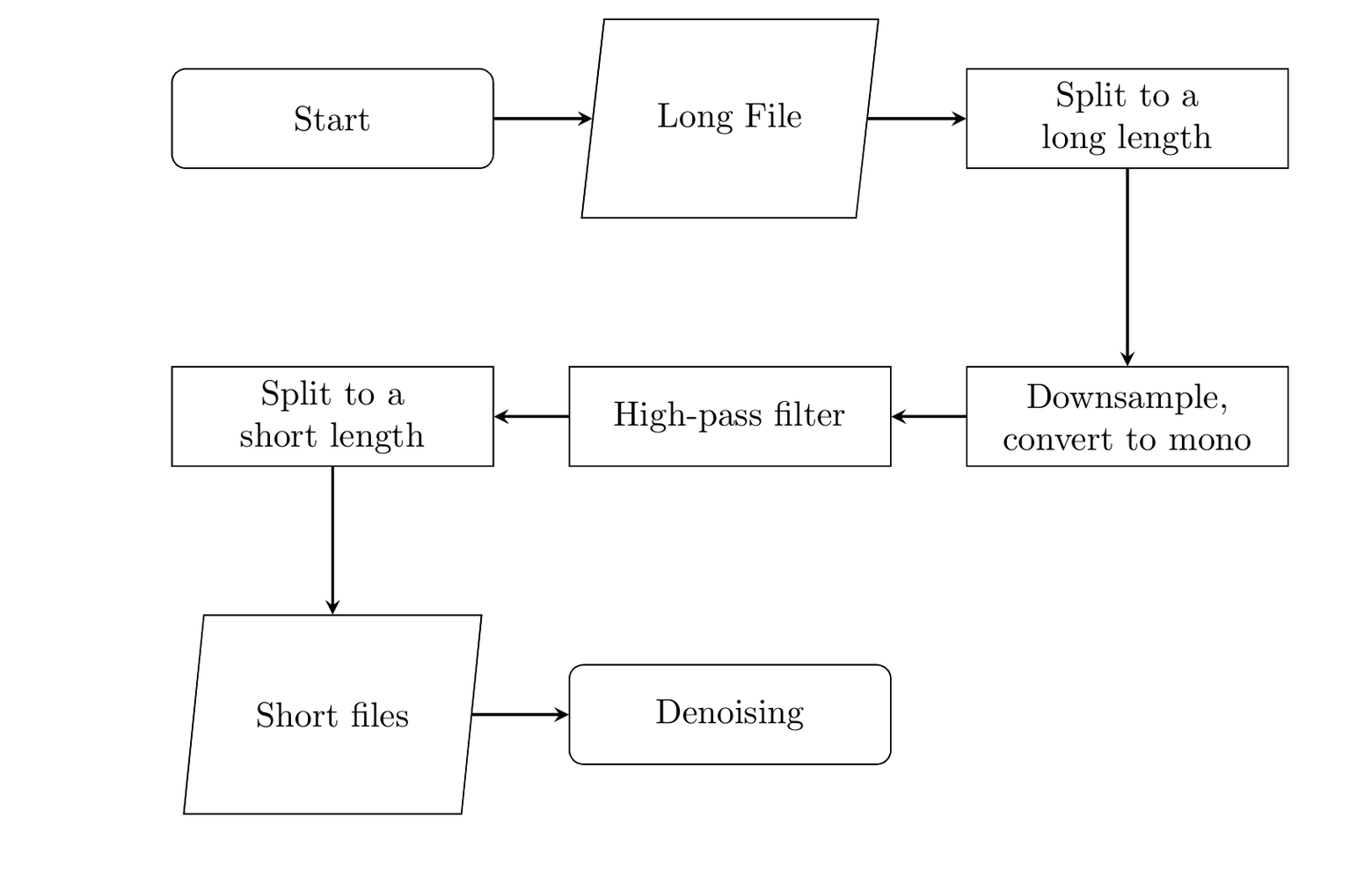}
		\caption{\textbf{Early steps of the processing pipeline.} The ``long length" and ``short length" are determined in subsequent tests.}
		\label{Fig.:splitterFlowChart}
	\end{figure}

\begin{figure}
		\centering
		\includegraphics{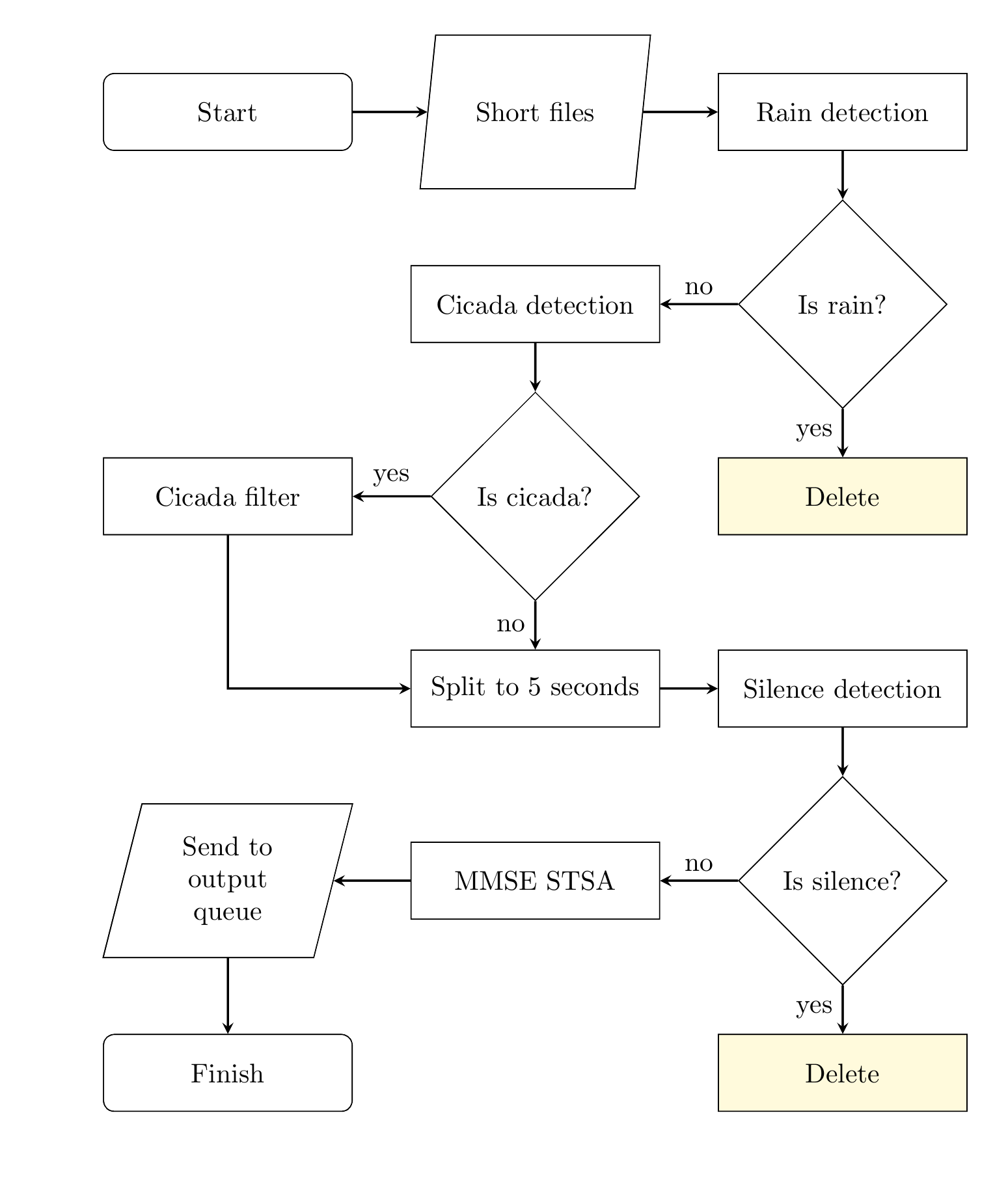}
		\caption{\textbf{Denoising steps of the processing pipeline}}
		\label{Fig.:denoiserFlowChart}
	\end{figure}

Based on the above findings and evaluation results from the previous sections, the final pipeline for preprocessing bioacoustics recording based on denoising filters is given in Algorithm~\ref{alg:pipeline} and summarised in Figs~\ref{Fig.:splitterFlowChart} and~\ref{Fig.:denoiserFlowChart}.

Files are first split to break up processing into smaller steps which can be parallelised. Compression processes are then applied to reduce execution time of all other processes. High-pass filtering is applied, removing any noise below 1 kHz and improving detection mechanisms. This also works better with longer split lengths, so applying earlier improves execution times. Then rain and cicada detection are executed, with rain detection executing earlier because it may eliminate audio from further processing. Files are then split to 5 seconds, before silence detection is performed. Finally, the MMSE STSA filter is executed. Placing this at the end reduces execution time because any files removed by other processes do not need to undergo MMSE STSA filtering, which has the longest execution time of any individual process.

Importantly, any file removed in earlier processes does not need to complete the pipeline, saving significant execution time. Hence, silence and rain detection steps significantly improve execution times, while resulting in higher quality output because useless chunks are discarded. In particular, skipping the MMSE STSA step removes the majority of processing time of any given file.

The next section takes this processing pipeline and distributes it over multiple machines to greatly further execution times.

\section*{Scalable distribution of the preprocessing pipeline}
\label{sec:distributedSystem}

This section describes the proposed approach for distributing work (i.e. the processing pipeline) amongst multiple machines, and evaluates this approach in terms of execution time, resource utilisation and load balancing. Results from these tests are used to improve the efficiency of the overall pipeline's execution for processing large recordings.

\subsection*{Master-slave system}
Our approach utilises a master-slave architecture with file parallelisation to progress through the processing pipeline. This architecture makes it easy to allocate work to slaves without the master needing to do much work itself. We constructed a bespoke master-slave system, as opposed to using an off-the-shelf approach, to avoid unnecessary overhead and to gain low level control over data flow. Files are processed through the pipeline on one slave each. This is chosen, as opposed to distributing work on a per-process basis, because workload can be evenly distributed among slaves by splitting files into small chunks. 

The master first splits, downsamples, and high-pass filters each file. The time taken to perform these steps is small compared to the overall processing time of the pipeline, so executing these steps in serial does not increase processing time. High-pass filtering is performed on the master process because it utilises long split lengths. By doing this on the master process, files can be split into shorter chunks for distribution. It then adds files it has processed into a queue. The master and slaves then communicate with each other about when they are ready to send and receive files. The master tracks which files have been sent to each slave, and which have completed processing, such that it can re-send files to different slaves if a slave disconnects or crashes.

Upon completing processing, slaves will send results back to the master. Results come in two forms: processed files and deleted files. If the slave sends a processed file, the name of the original file is first sent to the master, such that it can recognise that the file has been processed and the original file can be replaced, and then the processed files derived from the original file are sent. There are usually more processed files than original files, as files are split into 5 second chunks for silence detection. The functionality enabling slaves to send multiple files of different lengths for each file they received also allows for more flexibility as to how slaves process files in future work. In the case of samples identified for deletion, the slave simply sends the name of the file to delete and the master deletes its copy.

\subsection*{Slave parallelisation}

Parallelisation is performed both between multiple machines and between multicore processors. To parallelise work within a single machine, a central thread handles communication between the master and the slave, acting similarly to a secondary master (with threads being slaves). Files given to the slave from the master are added to a queue of files pending processing, which is managed by the central thread. The queue is set to a fixed size, such that if the queue falls below this size, the slave will request more files from the master. Processing threads then remove files from the queue and process them in the the denoising pipeline. Upon completing processing, results are sent to one of two queues managed by the central thread: one for processed files and another for deleted files. After a set time interval, all results are sent to the master and queues are cleared. 

Using a dedicated thread for communication allows processing threads to continually process audio without individually communicating with the master. This results in fewer requests to the master, reducing communication overhead.

\subsection*{Evaluation}

The approach for distributing the preprocessing pipeline described in the previous section is tested using several measures across several configurations to improve the system's scalability and to determine its time efficiency for preprocessing large recordings.

\subsubsection*{Methodology}

The testing methodology for this system is as follows:
\begin{itemize}
	\item Run a basic process in isolation that sends files from one machine to the other. Measure sending times with files of varying lengths (5--30 seconds), with 30 minutes (302 MB) of audio; repeat 5 times to observe variability
	\item Test the system by varying the following parameters:
	\begin{itemize}
	\item Split file length (5, 10, 15, 20, and 30 seconds, or 215, 430, 646, 861, and 1260 kB)
	\item Split file length prior to high-pass filtering (1--3 minutes, or 2.52--7.56 MB); hereafter referred to as the \textit{Long split length})
	\item Queue size of the central slave thread
	\item Frequency with which slaves send results
	\end{itemize}
	\item	Evaluation measures:
	\begin{itemize}
		\item Average processing time
		\item Average CPU and RAM usage for all machines
		\item Changes in execution time as slaves are added
		\item Load balancing
	\end{itemize}
	\end{itemize}

\subsubsection*{Communication times}
\begin{figure}
	\centering
	\includegraphics[width=0.8\linewidth]{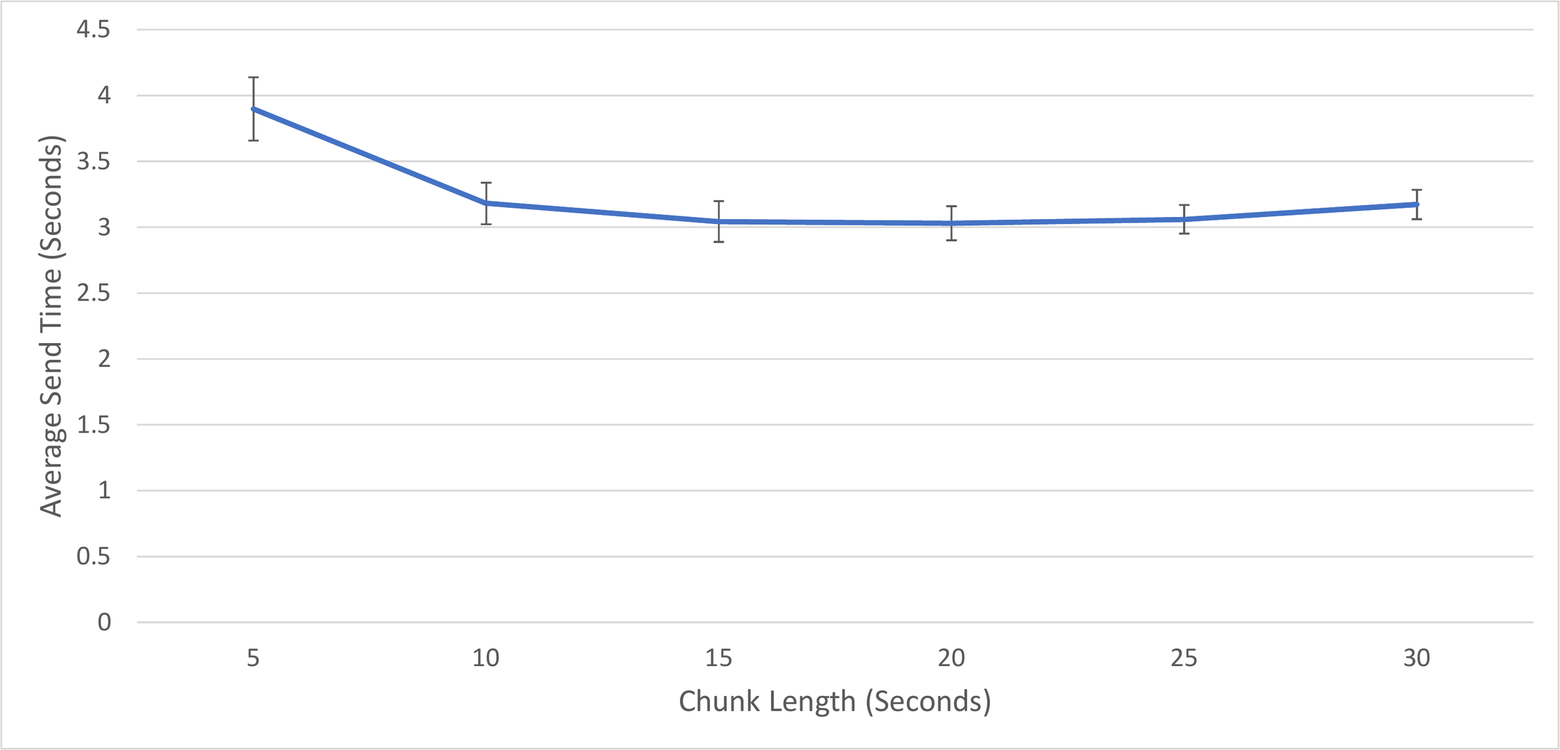}
	\caption{\textbf{File sending times} Time spent to send 30 minutes (302 MB) of audio back and forward between two virtual machines per split length}
	\label{Fig.:filesendtime}
\end{figure}

A short test was conducted where 30 minutes (302 MB) of audio, already split into short chunks of a fixed length, was sent back and forth between two machines, one chunk after another, with the aim of determining if file transmission took a significant amount of time, and if the sending time varies with split length. The total time taken to send all the files was recorded. The test was repeated five times for different file lengths. The results of this test are shown in Fig.~\ref{Fig.:filesendtime}.

This test shows that sending 5-second long chunks results in a slower sending time, whereas anything higher consumes about the same amount of time. Overall, the sending time is small relative to other computation, taking less than 4 seconds for all chunk sizes for 30 minutes (302 MB) of audio. This is equal to less than 16 seconds for two hours (1.2GB) of audio. This is a very small amount of time compared to the execution times of other processes in the pipeline, such as the MMSE STSA filter (see Fig.~\ref{Fig.:processtimesperprocess}). However, this becomes more significant as the number of processors increases because, while overall processing times are reduced, the communication time will remain approximately constant. 

Additionally, this is an idealised scenario in which files are sent and received in a predictable fashion. The distributed system used in processing the files is much more complicated, with slaves sending and receiving files as needed, creating a less predictable scenario. In a situation where multiple slaves are sending results or receiving files simultaneously, the sending time will inevitably increase.

Overall, this test shows that communication between the master and the slaves has a small, but not insignificant effect on overall processing time, although changing the split length could only give a 1 second saving per 30 minutes of audio at most, under ideal conditions. It is overall likely insignificant compared to other factors.

\subsubsection*{Identifying best settings for efficiency}
\label{sec:bestSettings}

A large number of configurations were examined to find which set produces the fastest execution. In particular, the amount of processing conducted by the master thread prior to sending to the slaves, the split length, the split length before applying the high-pass filter, referred to here as the \emph{long split length}, the maximum queue size of slaves' central threads, and the interval between slaves sending results are considered. These tests are carried out using 4 virtual machines with 4 cores each and 16 GB of RAM. These machines are hosted in the Nectar Cloud, which is a cloud platform used by Australian and New Zealand universities.

Initial ad hoc testing was conducted using a large number of different parameter sets to reduce the number of configurations to undergo more thorough testing to a more manageable level. In these tests, each set was only tested once. From this ad hoc testing, parameter ranges were set to evaluate 90 configurations in more depth. Each test was conducted five times each with the same two hours of audio used in earlier tests being processed each time. Of these, 10 configurations with the lowest average execution time are shown in Table~\ref{bestConfigs}.

	\begin{table*}
	\caption{Ten best configurations identified in distribution testing.}
	\begin{tabu} to 1\textwidth{X[1,c]X[1,c]X[1,c]X[1,c]X[1,c]X[1,c]} 
		\toprule
		Split length (s) & Long split length (s) & Max queue size & Time per send (s) & Average execution time (s) & Std. dev. (s) \\ 
		\midrule
		10&120&7&2&72.55&1.14 \\ 
		20&60&5&2&72.74&0.90 \\
		10&60&5&2&72.75&0.56 \\
		5&120&7&3&72.76&1.13 \\
		30&60&3&2&72.95&0.42\\
		10&120&5&3&72.95&0.45\\
		15&60&5&3&73.14&0.70\\
		5&60&7&4&73.14&1.41\\
		10&60&7&2&73.15&1.00\\
		20&60&3&2&73.15&1.58\\
		\bottomrule
	\end{tabu}
	\label{bestConfigs}
	\end{table*}

A key insight from these results is that there is little difference in performance between the best configurations, with the top 10 being separated by 0.6 seconds over 2 hours (1.2 GB) of audio (0.8\% of the fastest time) and well within the standard deviation of all the top 10. These equivalently process audio at a rate of $16.4\textup{--}16.5\pm0.4 \textrm{MB}s^{-1}$ (error given by the maximum standard deviation). The only poor combination found is to have a split length of 5 and maximum slave queue size of 3, and any combination of other settings. These configurations are about 25 seconds slower on average than any other configurations. The top 84 configurations (i.e. all configurations except the known bad ones) are separated by 8.03 seconds (this becomes 2.81 seconds for the top 50), which is statistically significant, so there is a small time efficiency advantage from thoroughly testing configurations as opposed to selecting one at random. 

This indicates that we can select configurations for accuracy, without significant loss of efficiency. Because splitting into 15 second chunks was the most accurate approach for removing rain and cicada sounds, this is taken to be the split length in further testing. This gets split into 5 second chunks for silence detection at a later point of the pipeline.

\subsubsection*{Scalability testing and analysis}
A further test was conducted to determine how scalable the system is. The system was tested using two hours of audio known to contain bird sound, rain, cicada choruses, and silence with varying numbers of machines. The test was run four times for each case, and the average execution time recorded. The 1-core execution test used a process specifically written for sequential execution, while the others used the distributed system. The CPU count includes the master and slave nodes. Because the master node does not require a large amount of resources, a slave node is also executed on the same machine as the master. Each instance tested contained 4 cores and 16 GB RAM, though most of this RAM is not used by the system. The 2-core case was tested using a single 2-core instance running a master and a slave process.

\begin{figure}
	\centering
	\includegraphics[width=0.8\linewidth]{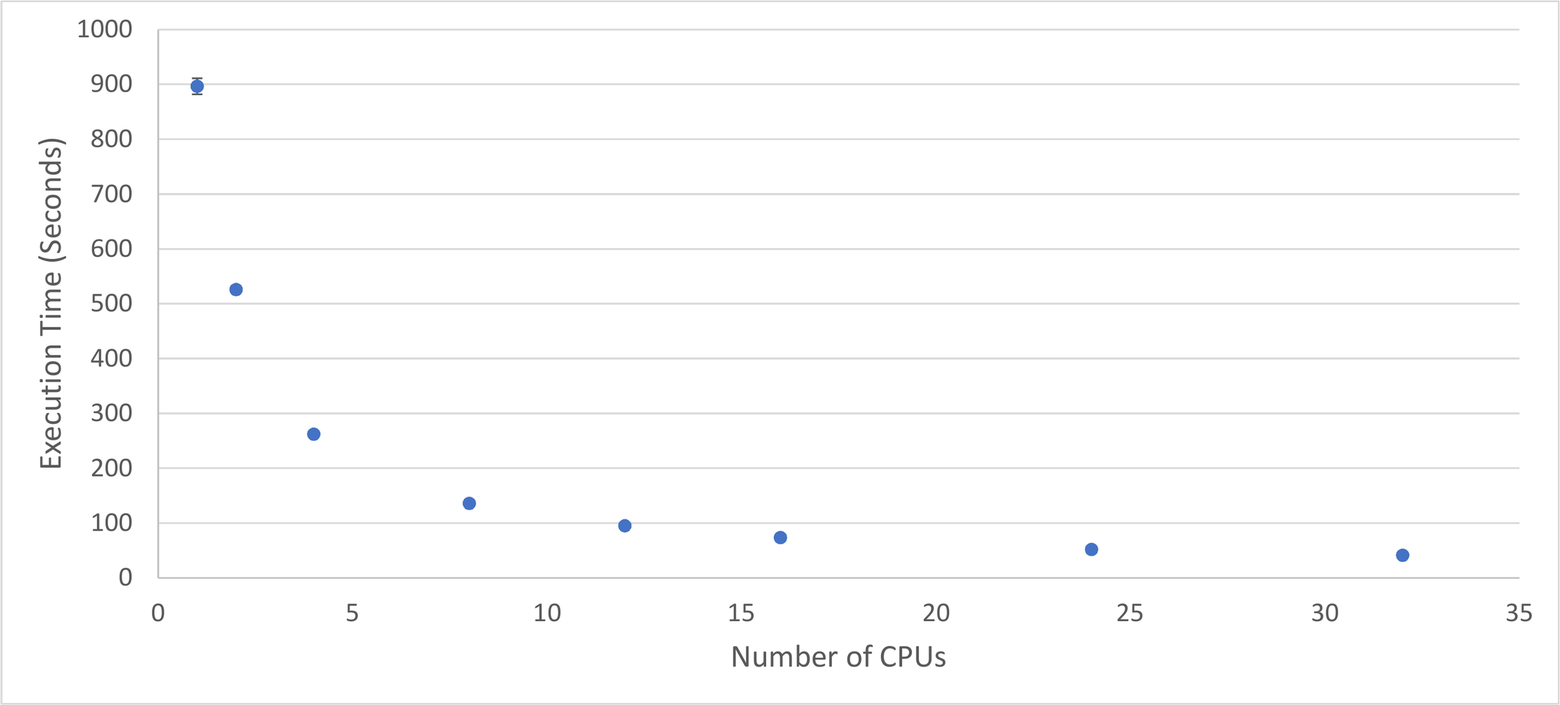}
	\caption{\textbf{Average Execution time of the system given a number of cores.} The master and each slave have 4 cores, so 16 cores uses 4 virtual machines. Standard deviations are too small (4.9 seconds at most) for error bars to be visible}
	\label{Fig.:scalableexectime}
\end{figure}

\begin{figure}
	\centering
	\includegraphics[width=0.8\linewidth]{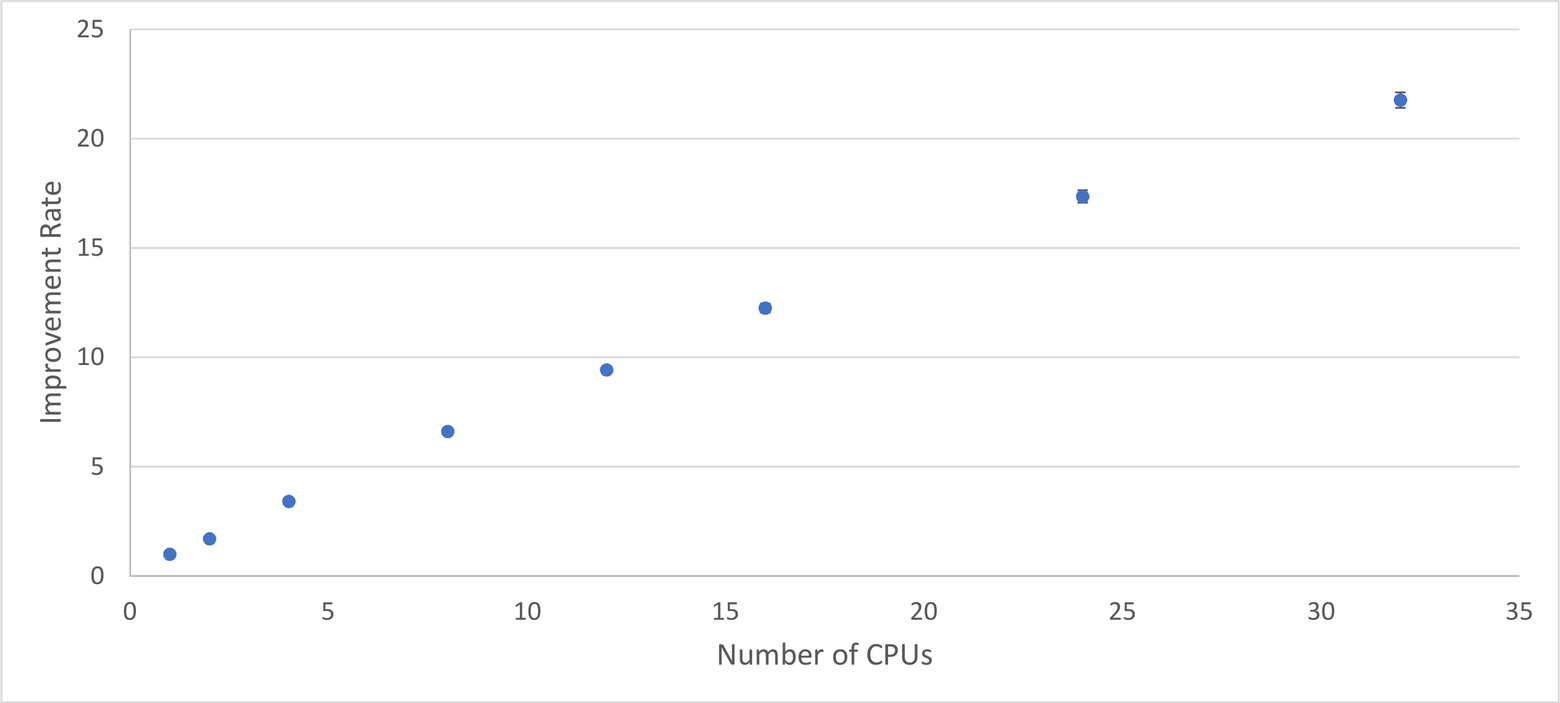}
	\caption{\textbf{Rate of improvement in execution time per number of cores.} This is given by Execution Time of 1 core/Execution time of $x$ cores.}
	\label{Fig.:scalableexectimenorm}
\end{figure}

Fig.~\ref{Fig.:scalableexectime} shows the average execution time for the number machines used. Fig.~\ref{Fig.:scalableexectimenorm} presents the improvement in the execution time over 1 core by measuring how many times faster execution is compared to the sequential (1-core) case.

These figures show that the system is indeed scaling almost linearly, with significant speed boosts from using extra processors. The improvement rate does begin to slightly diverge from perfect linearity when high numbers of cores are used, but even a 32-core distributed system still shows significant performance increases over a 24-core system. There is also a slight statistical anomaly where the 2-core system does not improve as much over the sequential 1-core system as might be expected. This is likely because of the extra overhead involved in using the distributed system over the sequential system. However, this extra overhead does not seem to prevent the system from being linearly scalable. 


%
%
%

A further test was conducted using smaller machines which when combined give a similar power level to large machines. The configurations compared are as follows:

\begin{enumerate}
\item One 4-core, 16 GB RAM master, one 4-core, 16 GB RAM slave
\item One 4-core, 16 GB RAM master, two 2-core, 6 GB RAM slaves
\item One 4-core, 16 GB RAM master, four 1-core, 4 GB RAM slaves
\end{enumerate}

The master also runs a slave instance in all cases, to make a fairer comparison with the previous tests. This also has the effect of testing system performance where different sizes of virtual machines are operating at the same time, as the master virtual machine runs a slave with 4 cores in all cases, albeit while competing for resources with the master thread.

\begin{figure}
	\centering
	\includegraphics[width=0.8\linewidth]{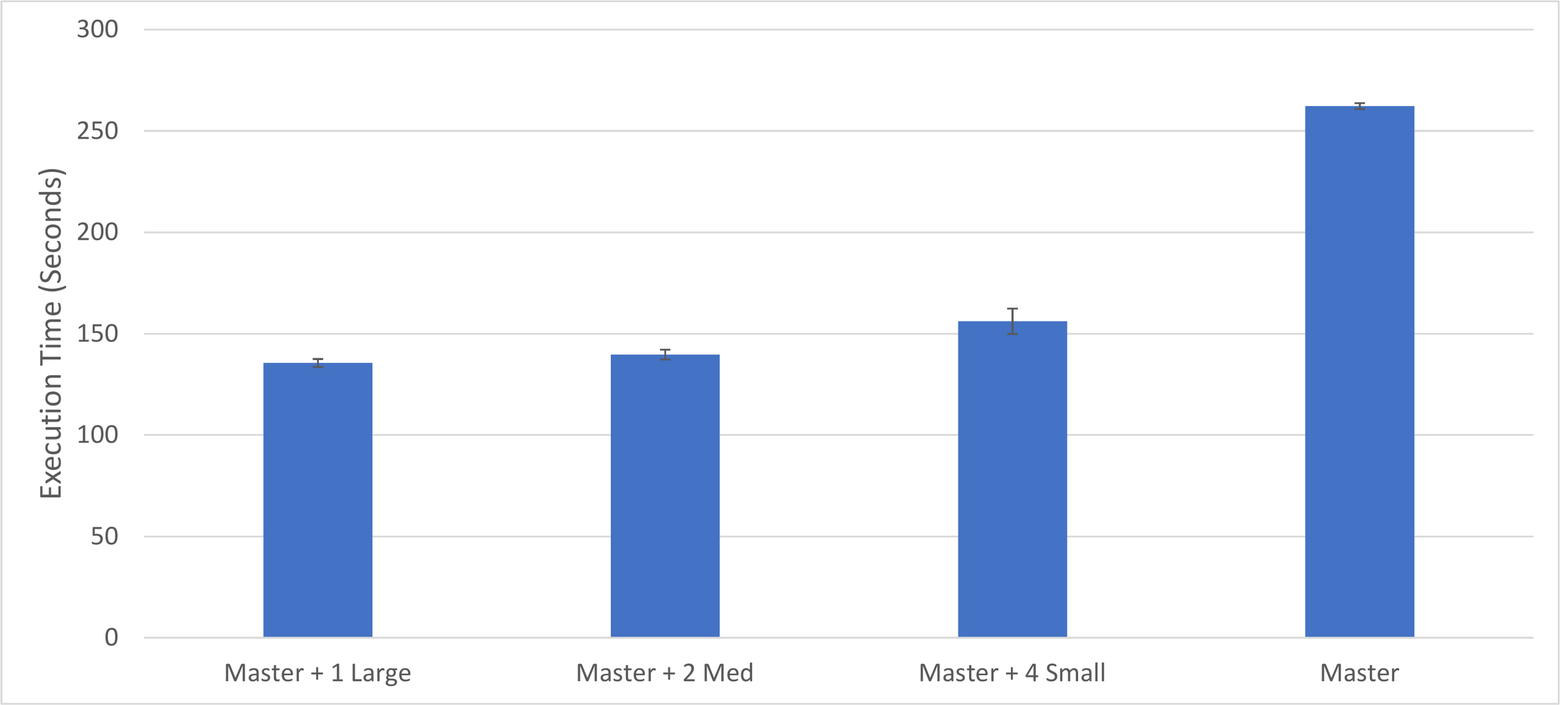}
	\caption{\textbf{Execution time comparison between using more smaller machines and using fewer larger machines.} The master on its own is also shown for comparison}
	\label{Fig.:scalableexectimeinstances}
\end{figure}

The results shown in Fig.~\ref{Fig.:scalableexectimeinstances} indicate that the system works as well with the master and two 2-core slaves compared to the master and one 4-core slave, and slightly worse when four 1-core slaves are used. The slower execution time when using 1-core machines could be due to extra overhead caused by the use of the centralised slave thread. This use of the central slave thread (which can be further broken down into six small threads) results in excessive overhead with smaller machines, while with larger machines reducing the amount of communication to the master and waiting times in processing files become advantageous. It could also be due to an inappropriate queue size being used for smaller machines, leading to imbalances in workload during later stages of execution. The system is developed for larger machines, so it makes intuitive sense that they would compute faster. Overall, the system is capable of performing efficiently with virtual machines of any size, although slightly less efficiently when 1-core machines are used. It also shows that the system can maintain efficiency when machines of different sizes are processing at once, because the master is running a slave thread with 4 available cores in all tests.

\subsubsection*{Load balancing testing and analysis}

An analysis of load balancing was also conducted at the same time as the scalability tests. This measured how many files are going to each of the slaves. Because all the slave machines have identical specifications, the file distribution should be even in an ideal case, outside of one slave which will have a lower number of files because it is sharing resources with the master process.

\begin{figure}
	\includegraphics{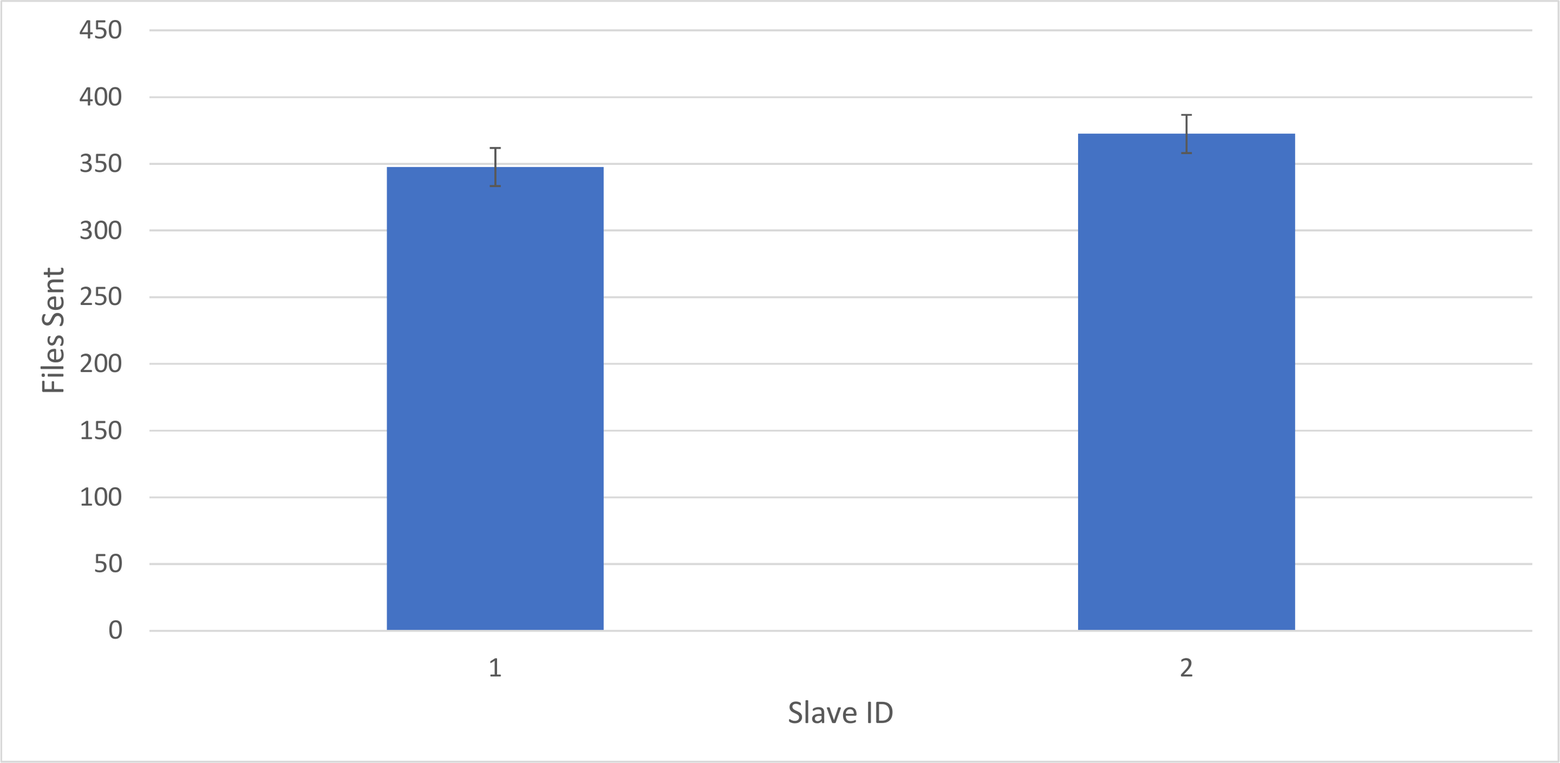}
	\caption{\textbf{Load distribution in processing for two slaves}. The number of files each slave processes is measured over four tests. Files are all of the same size.}
	\label{Fig.:2Slaves}
\end{figure}

\begin{figure}
	\includegraphics{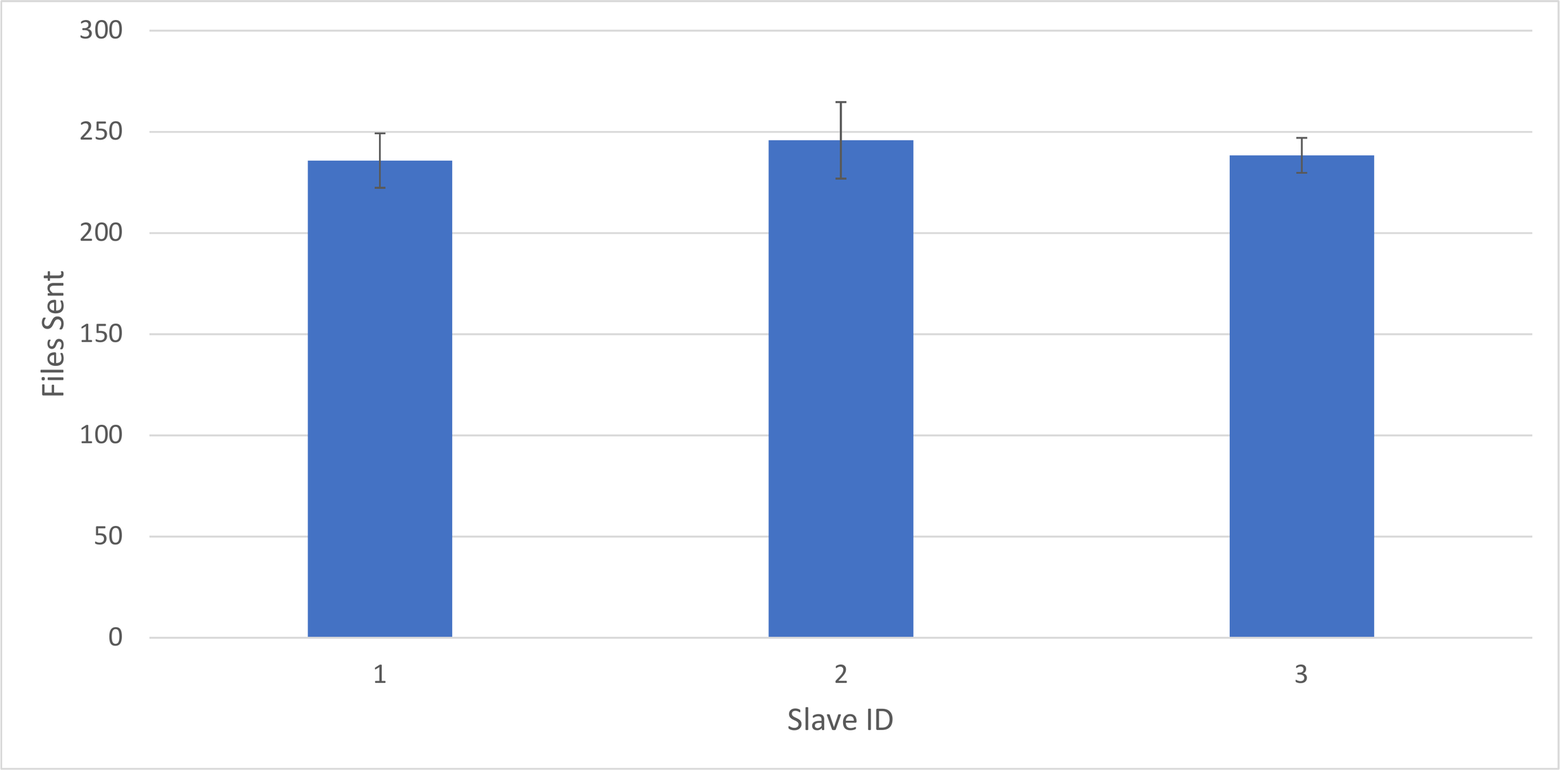}
	\caption{\textbf{Load distribution in processing for three slaves}. The number of files each slave processes is measured over four tests. Files are all of the same size.}
	\label{Fig.:3Slaves}
\end{figure}

\begin{figure}
	\includegraphics{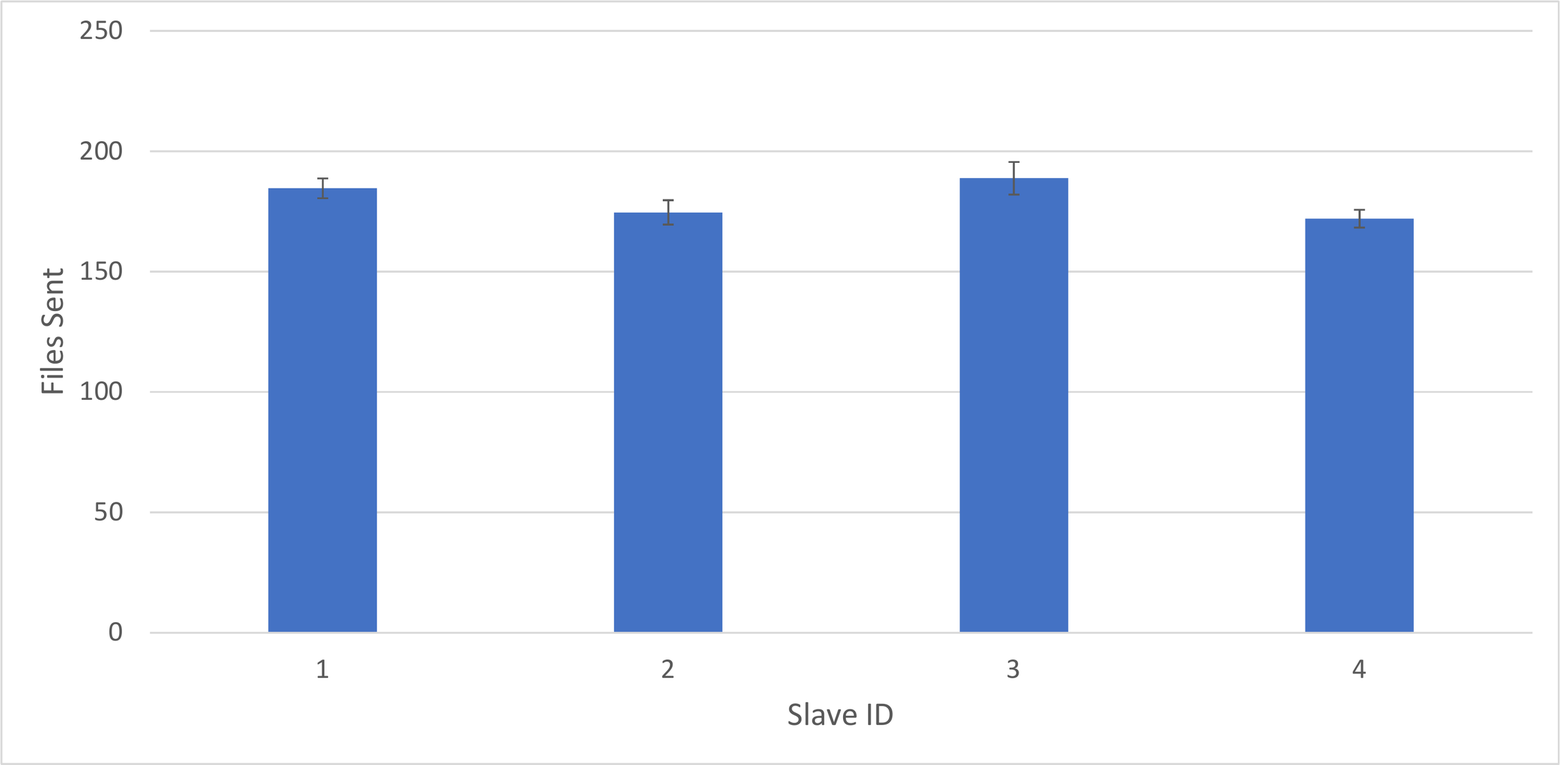}
	\caption{\textbf{Load distribution in processing for four slaves}. The number of files each slave processes is measured over four tests. Files are all of the same size.}
	\label{Fig.:4Slaves}
\end{figure}

Figs~\ref{Fig.:2Slaves} -- \ref{Fig.:4Slaves} show that the workload is well balanced, with each slave processing almost the same number of files in each test. This indicates that the system is distributing work evenly.


\begin{figure}
	\includegraphics{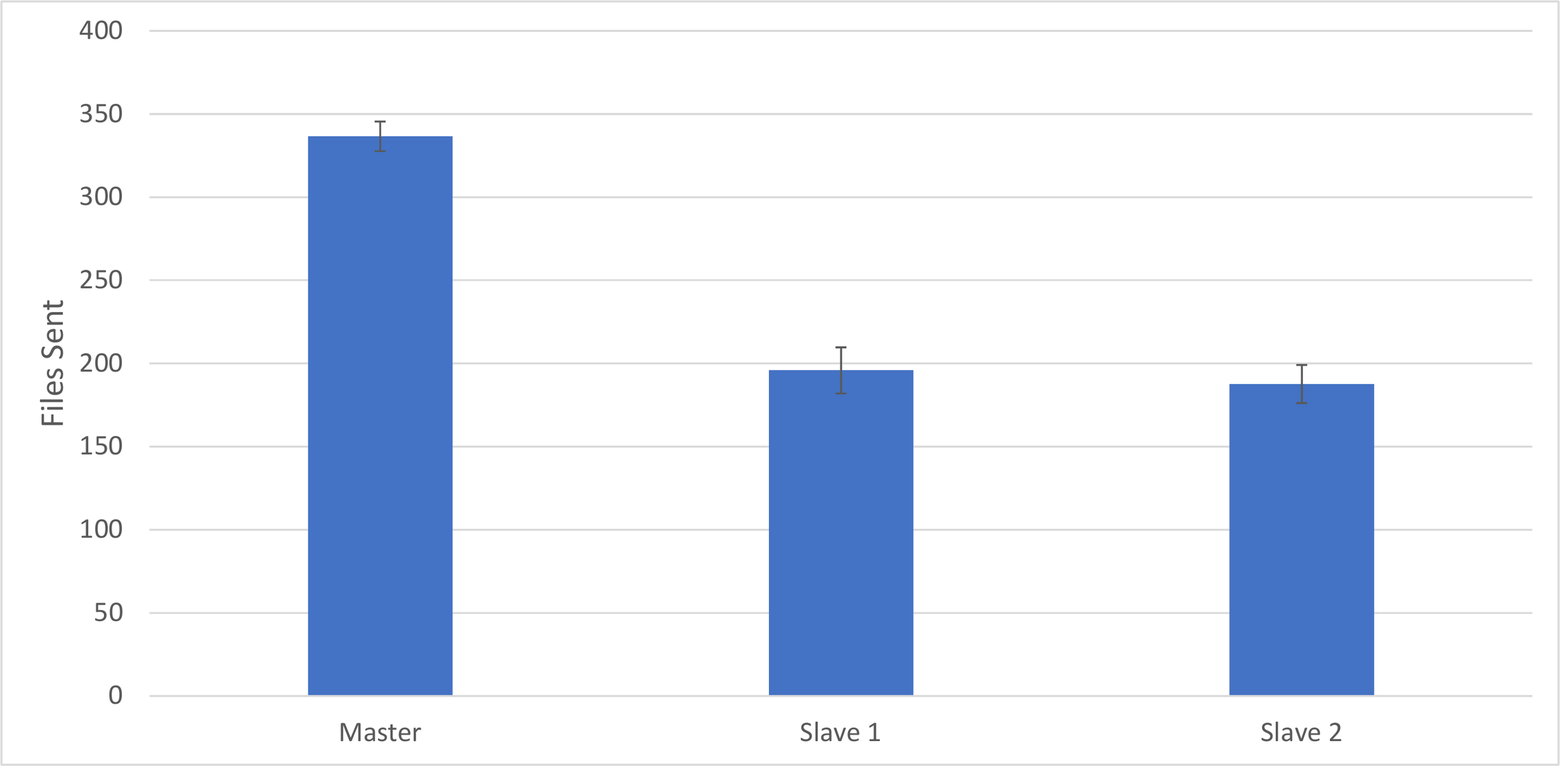}
	\caption{\textbf{Load averages between two 2 core slaves and one 4 core slave}. Load measured by the amount of files processed by each slave. The 4 core slave is also acting as a master}
	\label{Fig.:LoadBalance2Meds} 
\end{figure}

\begin{figure}
	\includegraphics{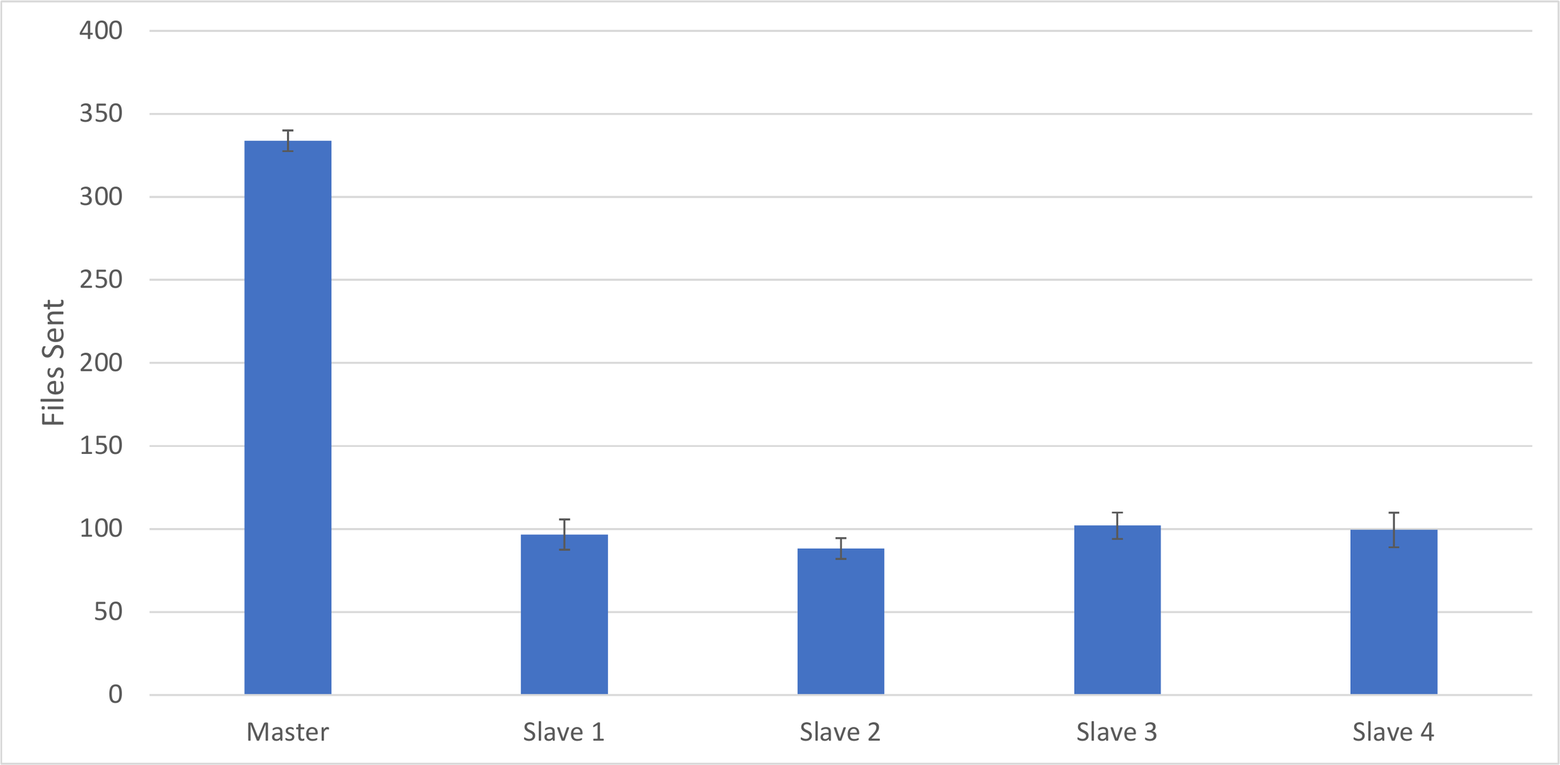}
	\caption{\textbf{Load averages between four 1 core slaves and one 4 core slave}. Load measured by the amount of files processed by each slave. The 4 core slave is also acting as a master} 
	\label{Fig.:LoadBalance4Meds}
\end{figure}

Figs~\ref{Fig.:LoadBalance2Meds} and \ref{Fig.:LoadBalance4Meds} demonstrate that the system is capable of balancing workload where the machines being used are of unequal power. This data are taken from earlier tests where the master, with 4 cores, is running a slave process simultaneously and less powerful machines are also running slave processes. Here, the master correctly allocates more files to itself compared to what it allocates to each of the slaves, proportional to the differences in computing power.

\subsubsection*{Resource usage test and analysis}

A test was conducted to see how efficiently the system is using resources. This was done by processing two hours of audio with four slaves, and sampling the CPU and RAM usage approximately every 8 seconds. This sampling was done using a shell script running in parallel to Java execution, although some data regarding timing is sent to the debugging logs to help synchronise the timings between slaves. While accuracy of the times is imperfect, it should be accurate to within 3 seconds.

\begin{figure}
	\centering
	\includegraphics[width=0.8\linewidth]{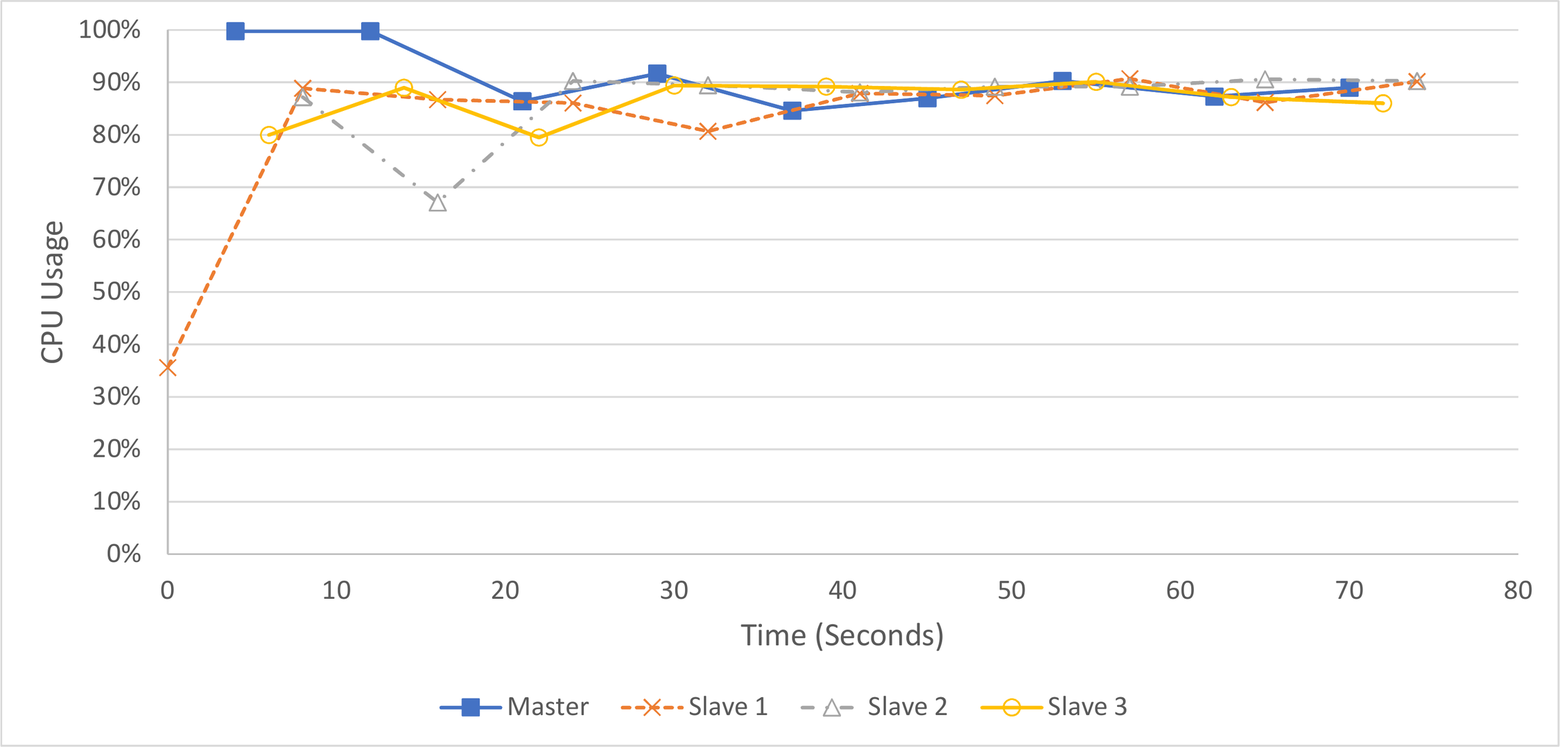}
	\caption{\textbf{CPU Usage over four 4-core machines processing 2 hours (1.2 GB) of audio}}
	\label{Fig.:cpuusage}
\end{figure}

Fig.~\ref{Fig.:cpuusage} shows that CPU usage remains at about 90\% for most of the processing of the two hours of audio. There does appear to be a slight drop below this number at the start of processing, presumably due to the master still performing early processing and not having files to send. Overall, assuming the overhead is not significant to CPU usage, it would be difficult to significantly improve upon the current pipeline without changing the pipeline itself. Note that the master is also running as a slave, and the master CPU usage relates to the usage by the slave and master processes running on that machine.

\begin{figure}
	\centering
	\includegraphics[width=0.8\linewidth]{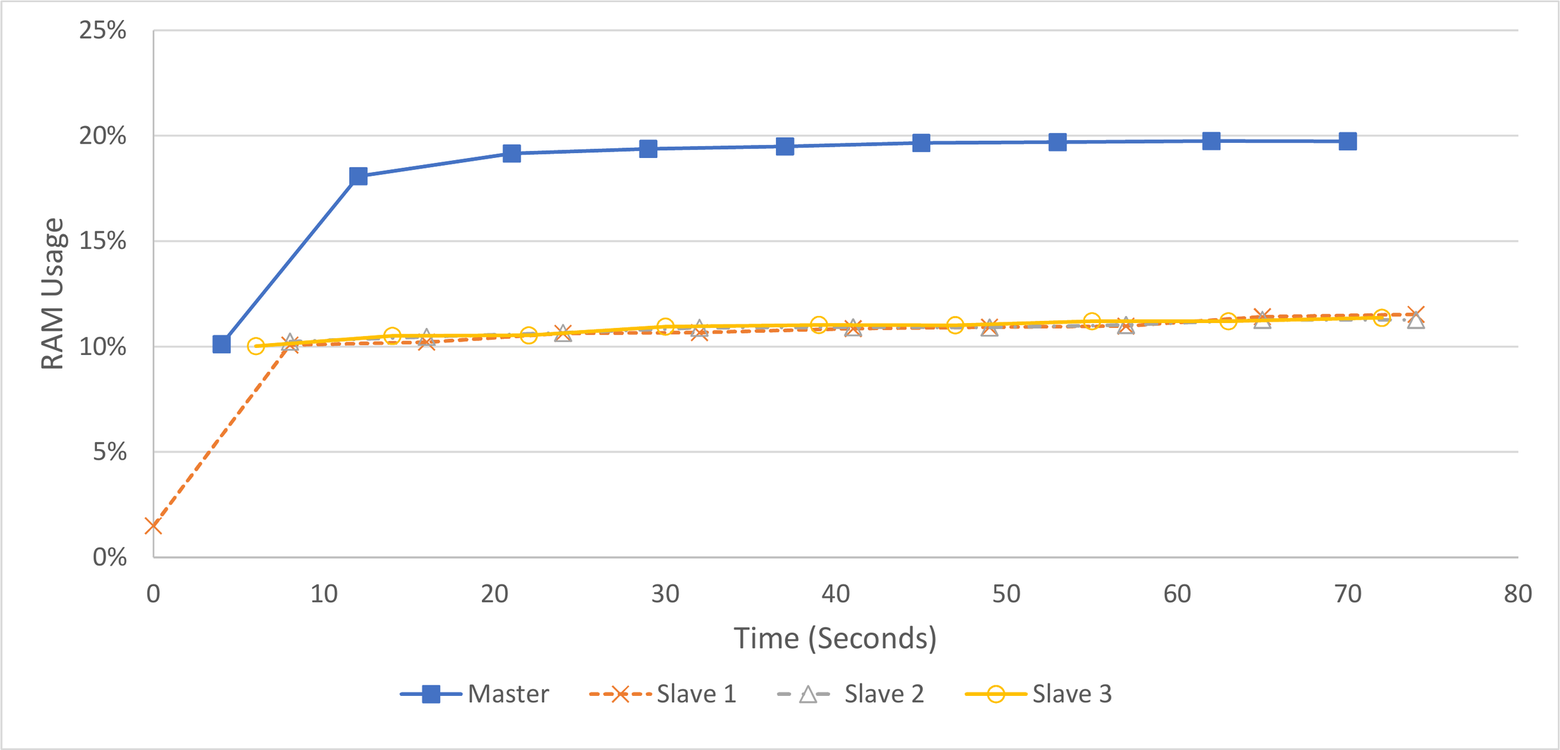}
	\caption{\textbf{RAM Usage over four 4 core Machines processing 2 hours (1.2 GB) of audio}}
	\label{Fig.:ramusage}
\end{figure}

Fig.~\ref{Fig.:ramusage} shows that the three slaves utilise around 11\% of the machines’ 16 GB of available RAM, remaining constant after the first 10 seconds. The master uses more RAM, presumably due to holding information about slave sockets and data streams, as well as information about files, relating to whether they have been sent and which slave is processing them, in addition to running a slave process.

RAM is underutilised overall. The system relies heavily on file writes and file reads using hard drives, which results in low RAM utilisation. Keeping more data in RAM could result in faster memory access and in turn, faster processing. However, as CPU usage is already fairly high, hard disk reading and writing does not seem to be a significant bottleneck in processing these audio files. Nonetheless, this is a potential area for performance improvements in future work.

\section*{Comparison with similar approaches}
\label{sec:relWork}

Dugan et al.\cite{Dugan_2011} focus their cloud infrastructure on completing two tasks: auto detection and noise analysis. In each of these, a process manager divides work into $M$ nodes which each independently work on their own tasks. Their sensor data is multiplexed in the data files (i.e. data from multiple sensors are shared in the one place), so data are divided by time, rather than by sensor. Recordings for the time period to be analysed are split into blocks equal to the number of processing nodes and each of these blocks are assigned a node. Nodes process independently, then return their output. Using this they found that, while speed improvements varied between the process being tested, the most improved process (classifier-based detection) was 6.57$\times$ faster for an 8-node server over a serial process, although another process (template-based detection) only improved by 3.33$\times$ over a serial process using an 8-node server running in parallel. A drawback to their approach is the use of a MATLAB package to handle distribution, which, while easier to develop, lacks low-level control over the data, and adds overhead. They have expanded this work with numerous publications, such as in a 2015 work\cite{Dugan_2015} where they built an Acoustic Data-Mining Accelerator (ADA), which parallelises mapping and gathering operators in an otherwise sequential process.

Truskinger et al.\cite{Truskinger_2014} aim to extract acoustic indices to visualise their bioacoustics data. To do this, they distribute work by splitting audio into smaller chunks, similarly to Dugan et al.\cite{Dugan_2011}. The research claims it is not feasible to process audio files any longer than two hours due to the high amounts of RAM required, so they use a specialised program called mp3splt to divide the audio into 1-minute long chunks. A master task creates a list of work items for work tasks to do. Each work task is given a different chunk of audio to analyse. The results of these tasks are aggregated by the master task. Through this parallelisation, the execution time of an analysis task involving the computation of spectral indices is improved by a modest 24.00$\times$ for a 5 instance, 32 thread (with 32 cores per instance) distributed cluster over a single threaded process. While certainly an improvement, the parallelisation appears inefficient as the improvement rate is much lower than the increase in resources. While discussion of the pipeline is not detailed in the paper, a possible reason for this low improvement rate is that there is a large serial component to the processing pipeline used and so the parallel processors are not fully utilised.

Thudumu et al.\cite{Thudumu_2016} developed a scalable framework to process large amounts of bioacoustics data using Apache Spark Streaming\cite{SparkStreaming} and the Hadoop Distributed File System (HDFS)\cite{Shvachko_2010} which utilised a master-slave model. The system parallelises the chunking of audio data and the generation of spectrograms. Parallelisation is handled by Hadoop and Spark. For a task involving splitting 1 GB of audio into 10 second chunks and generating spectrograms, the system showed a 4.50$\times$ improvement in execution time in a test with a 1 core master node and a 4 core slave node, but a weaker 7.50$\times$ improvement in execution time with a 1 core master and three 4 core slaves compared to a serial process, indicating the system is not as scalable as it could be. Using an equivalent number of processing resources, our system achieves a 9.98$\times$ improvement, with a much more computationally intensive processing pipeline.

\section*{Conclusions and future directions}
\label{sec:conclusion}

In this work, we derived an approach for preprocessing high volume bird acoustic data quickly and efficiently. We achieved this by deriving a processing pipeline based on examining the processing time and accuracy of individual preprocessing tasks, and how these changed depending on how the audio is split into smaller chunks.

In testing individual components of the system, we found that the MMSE STSA filter consumes a very large amount of the execution time, meaning this should be executed as late as possible. We also found that high-pass and cicada filtering using SoX consumes more time when more, shorter files are being processed compared to fewer, longer files, which gave rise to an efficiency improvement.

From this individual component test, a processing pipeline is derived, and then applied in a distributed architecture, capable of processing on many machines at once. The resulting system is found to scale almost linearly, even when using 32 cores, which improved execution time by 21.76 times over serial processing. This compares favourably to existing research. It is also found that the system balances load evenly between machines, and can proportionally distribute more files to more powerful machines. Cores on all machines are found to consistently utilise 90\% of their available power, though RAM is underutilised.



While this work presents a strong basis for creating a fast, efficient, and scalable bird acoustic preprocessing pipeline, there is great potential for expansion in the future. 
 Silence detection currently performs poorly and is limited in that it can only choose to keep or drop 5-second long chunks. This is not a large problem for the present investigation, as we are more concerned with the efficient processing of data. However, if we wanted to improve the accuracy and utility of our pipeline, we could replace our relatively simplistic approach with one of many existing segmentation processes, which divide animal calls into syllables, often being insensitive to noise\cite{Ramli_2016,Zhang_2015}.


This processing pipeline is simple and generic enough such that additional noise reduction techniques could be added to the pipeline without difficulty. Adding additional processes to the pipeline would likely mean nothing more than inserting a new process in between two existing ones. Although this work focuses on the removal of noise from two sources, cicada choruses and rain, there are many other noise sources that could be targeted.



\bibliographystyle{ieeetr}
\bibliography{bibfile}

\end{document}